\documentclass[prb,twocolumn,superscriptaddress]{revtex4-1}

\usepackage{amsmath,amssymb}
\usepackage[dvipdfmx]{graphicx}
\usepackage{bm}
\usepackage{bbm}
\usepackage[hypertex,colorlinks,bookmarks=false,citecolor=blue,linkcolor=black,urlcolor=red]{hyperref}
\usepackage{color} 
\usepackage{srcltx} 
\usepackage{newtxtext,newtxmath}

\begin{document}

\title{Weyl superconductor phases in a Weyl-semimetal/superconductor multilayer}
\date{\today}
\author{Ryota Nakai}
\affiliation{Institute for Materials Research,
Tohoku University, Sendai 980-8577, Japan}
\email{r.nakai@imr.tohoku.ac.jp}
\author{Kentaro Nomura}
\affiliation{Institute for Materials Research,
Tohoku University, Sendai 980-8577, Japan}
\affiliation{Center for Spintronics Research Network, 
Tohoku University, Sendai 980-8577, Japan}

\begin{abstract}
Topologically nontrivial superconducting phases have been engineered in topological materials by the proximity effect in contact with conventional superconductors.
In this paper, by using the method of the Kronig-Penney model, we study the superconducting proximity effect in the bulk electronic states of Weyl semimetals by considering a multilayer structure consisting of Weyl-semimetal and superconductor layers.
Due to the proximity effect, two Weyl nodes are decoupled into four nodes of Majorana fermions resulting in Weyl-superconductor phases or three-dimensional extension of topological-superconductor phases. We find that mismatch of the Fermi velocity and potential barriers at the interface gap out Majorana nodes, thus turn Weyl-superconductor phases with four Majorana nodes into Weyl-superconductor phases with half of Majorana nodes and topological-superconductor phases with odd integer Chern numbers.
\end{abstract}

\maketitle

\tableofcontents

\section{Introduction}
\label{sec:introduction}

Topology appears in condensed matter physics through the geometric (Berry) phase of electronic states in the momentum space.
A group of materials characterized by a nontrivial topology are dubbed as topological materials.
Topological insulators have full-gap bulk states and protected metallic boundary states \cite{hasan10, qi11}. Topological superconductors have full-gap superconducting bulk states and protected gapless boundary states of a charge neutral fermion, referred to as the Majorana fermion \cite{sato17}.
Topological semimetals are another class of topological materials in which the conduction and valence bands touch at points or lines in the Brillouin zone \cite{murakami07,wan11,yang11,burkov11,xu11,armitage18}. The band-touching points in Weyl semimetals, the Weyl nodes, can be present when either time-reversal or inversion symmetry is broken, and they appear in pairs with the opposite chirality. Each node can be regarded as the Dirac monopole of the Berry curvature, where the total flux of the Berry curvature emanating from a node is an integer multiple of $2\pi$ depending on the chirality \cite{berry84,fang03,turner13}. The presence of the Weyl nodes is related to transport phenomena, such as the anomalous Hall effect and the negative magnetoresistance \cite{yang11,burkov11,xu11,nielsen83,son13,burkov15,huang15,lundgren14,lucas16,spivak16} via the chiral anomaly \cite{zyuzin12,goswami13}.
The surface of Weyl semimetals hosts a gapless mode, the zero-energy states of which connecting two nodes with the opposite chirality are called the Fermi arc.
So far, Weyl semimetals have been realized in materials such as transition-metal monopnictides (inversion-symmetry-broken Weyl semimetals) \cite{huang15_2,weng15,xu15,lv15}, layered transition-metal dichalcogenides (type-II Weyl semimetals) \cite{soluyanov15,sun15,wang16-1} and Co-based magnetic Heusler compounds (time-reversal-symmetry-broken, or magnetic, Weyl semimetals) \cite{chang16,wang16-2,liu18,xu18,wang18}.

The Berry curvature can have a monopole structure even in superconductors and superfluids, the resulting phases of which are referred to as Weyl-superconductor/superfluid phases \cite{volovik87,murakami03,meng12,*meng17e,yang14,schnyder15,hermanns15}. The Weyl-superfluid phase has been studied in the context of the ABM phase of superfluid $^3$He \cite{volovik03}. The pair potential has pairs of nodal points at the position of monopoles in the momentum space where the superconducting energy gap closes.
Weyl superconductors have a Majorana arc on their surface at zero energy, and show anomalous thermal Hall effect depending on the position of nodes \cite{meng12,hermanns15}, analogous to the anomalous Hall effect in Weyl semimetals. 
In addition, Landau-level formation and chiral-anomaly-related phenomena have been studied in Weyl superconductors under strain \cite{massarelli17,liu17,matsushita18} and in the presence of the vortex lattice \cite{pacholski18}.

\begin{figure}
 \centering
 \includegraphics[width=70mm]{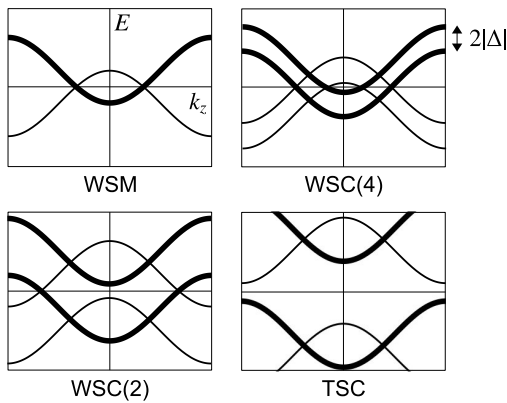}
 \caption{The band structure of a Weyl semimetal with an intrinsic superconducting order is shown in schematic pictures, which have been originally presented in \cite{meng12}. As the intrinsic pair potential $|\Delta|$ increases, a Weyl semimetal (WSM) is changed to a Weyl superconductor with 4 nodes [WSC(4)], then to that with 2 nodes [WSC(2)], and finally to a fully-gapped topological superconductor (TSC) characterized by a nontrivial Chern number.
 \label{fig:wsmwsc}}
\end{figure}

When a conventional superconducting order is induced in a Weyl semimetal, it can turn a Weyl semimetal into a Weyl superconductor \cite{meng12}.
To be more specific, consider a magnetic Weyl-semimetal model in the Nambu space with an intrinsic superconducting order given by, 
\begin{align}
 H(\bm{k})
 =
 \begin{pmatrix}
  h(\bm{k}) & i\Delta\sigma^y \\
  -i\Delta^{\ast}\sigma^y & -h^{\ast}(-\bm{k})
 \end{pmatrix},
 \label{eq:weylsemimetal_sc}
\end{align}
where
\begin{align}
 h(\bm{k})
 =
 \frac{k_z^2-K_0^2}{2m}\sigma^z
 +
 \lambda(k_y\sigma^x-k_x\sigma^y),
\end{align}
and $\sigma$ is the Pauli matrix of the spin degrees of freedom.
The eigenenergies of this model are given by
\begin{align}
 E^2
 =
 \left(
  \frac{k_z^2-K_0^2}{2m}
  \pm
  |\Delta|
 \right)^2
 +
 \lambda^2(k_x^2+k_y^2).
 \label{eq:modelwsmsc_eigenenergy}
\end{align}
Since, when $k_x=k_y=0$, the spin-up ($\sigma^z=1$) electronic band and the spin-down ($\sigma^z=-1$) hole band have the same dispersion along $k_z$, the superconducting order $\Delta$ lifts the degeneracy of these bands as shown in Fig.~\ref{fig:wsmwsc}, and thus two nodes of the Weyl fermions are split into four nodes of the Majorana fermions. Further increasing the pair potential, Majorana nodes are gapped out in pairs, and finally a topological superconductor characterized by the Chern number in the $k_x$-$k_y$ space is realized.
However, this scenario of the intrinsic superconductivity would not be feasible due to vanishingly small electronic density of states in Weyl semimetals near the node.

Topologically nontrivial superconducting phases have been tuned and engineered in strong spin-orbit-coupled materials by making contact with conventional superconductors.
When the surface of a topological insulator is attached with a conventional superconductor, the gapless surface mode turns into a fully-gapped two-dimensional topological superconductor due to the superconducting proximity effect \cite{fu08,akhmerov09}.
Even a normal metal can become a topological superconductor when Rashba spin-orbit coupling and the proximity effect are present \cite{lee09,sau10,alicea10}.
The surface of an inversion-symmetry-broken Weyl semimetal hosts a flat band of the Majorana fermions by attaching superconductors with $\pi$-phase difference making a Josephson junction \cite{chen16}.
However, the chiral surface state of a time-reversal-symmetry-broken Weyl semimetal is robust against the proximity effect, which lifts the degeneracy of the electron and hole bands, but does not gap out the surface state. \cite{khanna14}.
In the case of two dimensions, a quantum-anomalous-Hall insulator in proximity to a conventional superconductor realizes a topological-superconductor phase \cite{qi10}.

By stacking two different materials repeatedly, a multilayer model of an normal insulator and a three-dimensional magnetically doped topological insulator (TI-NI) has been invented to realize Weyl-semimetal phases \cite{burkov11}.
When topological-insulator layers are replaced by Weyl-semimetal layers (WSM-NI), quantum-anomalous-Hall-insulator phases with arbitrary Chern numbers appear, where Weyl-semimetal phases continuously connect quantum-anomalous-Hall-insulator phases with different Chern numbers \cite{yokomizo17}.
On the other hand, when normal-insulator layers are replaced by superconductor layers (TI-SC), the proximity-induced superconducting order splits two Weyl nodes of the TI-NI model into four Majorana nodes, and Weyl-superconductor phases are realized when the phase of superconducting layers is the alternation of 0 and $\pi$ \cite{meng12,*meng17e}.
From a different perspective, a multilayer model comprising alternating layers of even- and odd-parity orbitals has also been studied  \cite{das13}. This model realizes Weyl-semimetal and Weyl-superconductor phases in the presence of spin-orbit coupling and a chiral-$p$-wave superconducting order, respectively.

In this paper, we investigate the superconducting proximity effect in the bulk electronic states of a time-reversal-symmetry-broken Weyl semimetal by considering a multilayer structure of the alternation of Weyl-semimetal and superconductor layers (WSM-SC).
Thus, our model is a hybrid of the WSM-NI model \cite{yokomizo17} and the TI-SC model \cite{meng12}, and inherits physical properties from both models, that is, our model realizes both Weyl-superconductor phases and topological-superconductor phases with arbitrary Chern numbers.
In addition, our model has controlable transparency of the interface by inserting potential barriers between layers, and does not require $\pi$-phase difference of superconductor layers like in the TI-SC model \cite{meng12,*meng17e}.
Here, we notice that, in our model, the superconducting pair potential and the exchange field are given \textit{a priori}, homogeneous within each layer, and common for all layers, respectively, although the superconducting order can also be studied in a Kronig-Penney-type superlattice model \cite{tanaka89,tanaka91}.
In addition, the exchange field is assumed to be perpendicular to the interface in order to address the electronic states analytically, and thus the chirality blockade \cite{bovenzi17,dutta19} does not occur.

This paper is organized as follows.
In Sec.~\ref{sec:model}, we define a multilayer model studied in this paper.
Using this model, first, we study the case when the Fermi level of the Weyl semimetal lies at the node in Sec.~\ref{sec:electronicproperties}. The phase diagram without potential barriers is given in addition to the position of nodes, and the low-energy properties near a node is shown to be described by the Weyl Hamiltonian by using the degenerate perturbation theory. 
Then, we show that each phase is characterized by quantized or continuously varying thermal Hall conductivity, and finally the phase diagram with potential barriers is presented. 
In Sec.~\ref{sec:awayfromthenode}, we proceed to the study of the case when the Fermi level of the Weyl semimetal is away from the node.
In Sec.~\ref{sec:conclusion}, we summarize the results.

\section{Multilayer model}
\label{sec:model}

\begin{figure}
 \centering
 \includegraphics[width=30mm]{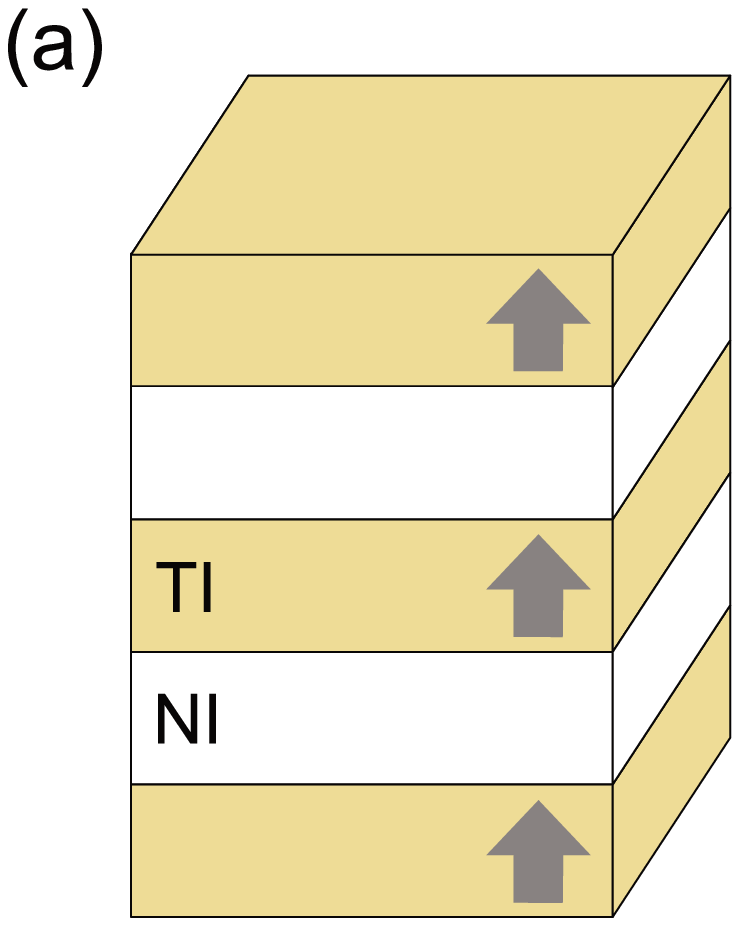}
 \includegraphics[width=30mm]{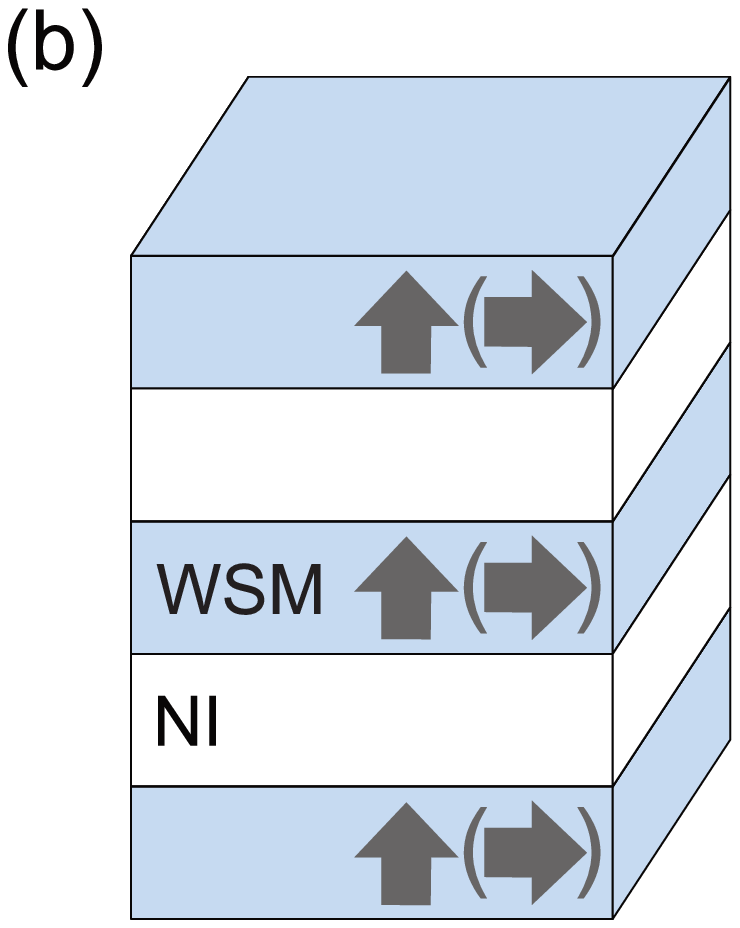}
 \includegraphics[width=30mm]{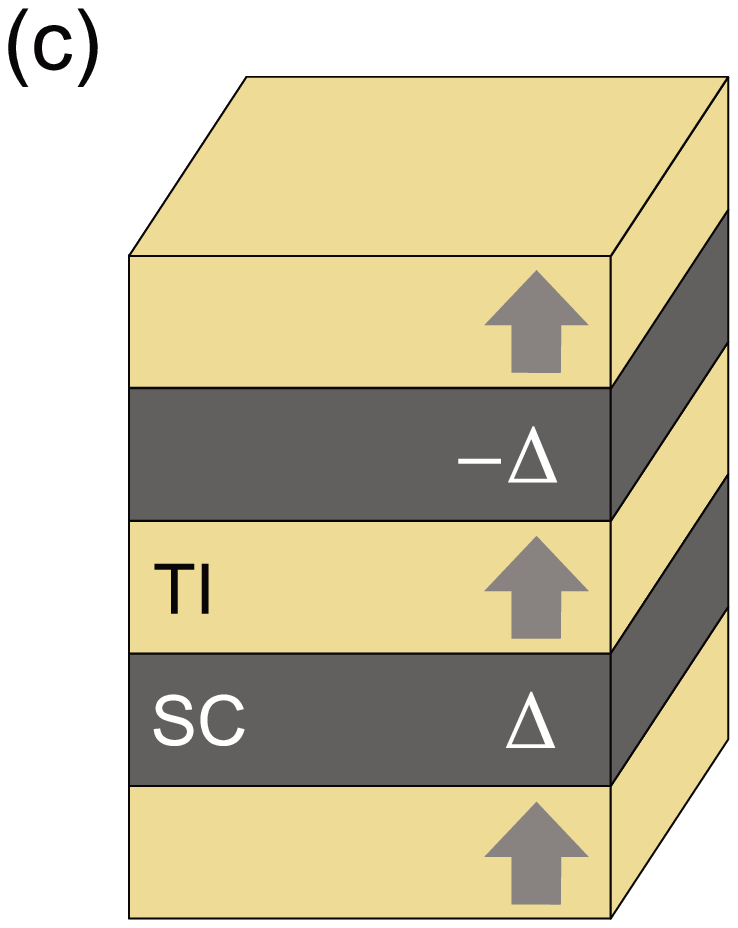}
 \includegraphics[width=30mm]{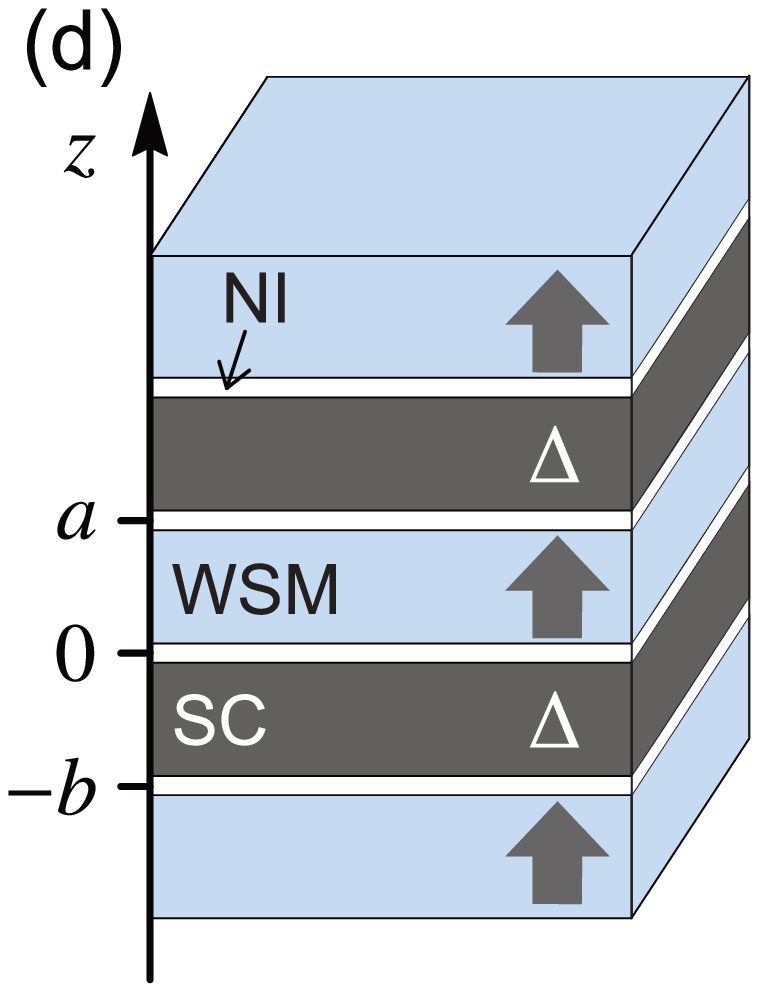}
 \caption{The multilayer models that consist of the alternation of a topological-insulator (TI)/Weyl-semimetal (WSM) layer coupled with an exchange field and a normal-insulator (NI)/superconductor (SC) layer are shown. In contrast to (a) the TI-NI multilayer model, our model [(d)] is a hybrid of (b) the WSM-NI model and (c) the TI-SC model. In addition, potential barriers, which work as a normal-insulator layer, between WSM and SC layers are contained in our model.
 \label{fig:multilayer}}
\end{figure}

We consider an alternating layered structure consisting of a time-reversal-symmetry-broken Weyl semimetal and an s-wave superconductor as shown in Fig.~\ref{fig:multilayer}(d).
We assume that time-reversal symmetry is broken in the Weyl-semimetal layer by coupling with the exchange field, and that the pair potential $\Delta$ in the superconductor layer is given \textit{a priori} and is present only in the superconductor layers.
We solve this model in the same manner as the Kronig-Penney model \cite{yokomizo17}, which has been used to study the electronic band structure under a periodic potential in a crystal.

The unit cell consists of a Weyl-semimetal layer of the thickness $a$ and a superconductor layer of the thickness $b$, and layers are stacked in the $z$ direction [Fig.~\ref{fig:multilayer}(d)]. In addition, each interface has a $\delta$-function potential barrier with the height $h$, which works as an insulator layer inserted between Weyl-semimetal and superconductor layers.
The Hamiltonian of this system within a unit cell $z\in(-b,a]$ is given by
\begin{align}
 &
 H_\text{total}(k_x,k_y,z)
 =
 H(k_x,k_y,-i\partial_z) 
 +
 h\tau^z
 \left[
 \delta(z)
 +
 \delta(z-a)
 \right],
\end{align}
where
\begin{align}
 H(k_x,k_y,-i\partial_z)
 =
 \left\{
 \begin{array}{lr}
  H_\text{WSM}(k_x,k_y,-i\partial_z) \,\,& (z\in(0,a))\\
  H_\text{SC}(k_x,k_y,-i\partial_z) \,\,& (z\in(-b,0))
 \end{array}
 \right.,
\end{align}
and $\tau$ is the Pauli matrix for the Nambu space.
This model is solved by constructing a wave function that has an eigenenergy $E$ and momenta $k_x,k_y$ in the in-plane directions.
A Weyl-semimetal layer lies in $z\in(0,a)$ and the Hamiltonian is given in the Nambu space representing electron and hole degrees of freedom by
\begin{align}
 H_\text{WSM}(k_x,k_y,-i\partial_z)
 =
 \begin{pmatrix}
  h(k_x,k_y,-i\partial_z) & 0 \\
  0 & -h^{\ast}(-k_x,-k_y,i\partial_z)
 \end{pmatrix},
 \label{eq:hamiltonianweylsemimetalnambu}
\end{align}
and its electronic part by 
\begin{align}
 h(k_x,k_y,-i\partial_z)
 =
 \frac{-\partial_z^2-K_0^2}{2m}\sigma^z
 -
 \mu_\text{W}
 +
 \lambda(k_y\sigma^x-k_x\sigma^y),
 \label{eq:hamiltonianweylelectron}
\end{align}
where $\sigma$ is the Pauli matrix of the spin degrees of freedom. Throughout this paper, we use the unit $\hbar=1$.
The momentum $K_0$ represents the position of the Weyl nodes ($\bm{k}=(0,0,\pm K_0)$) in the bulk of the Weyl semimetal and is determined by the exchange field.
$\lambda$ is the magnitude of Rashba spin-orbit coupling. 
The eigenfunction for an eigenenergy $E$ is, in general, given by
\begin{align}
 \psi(z)
 =&
 \sum_{\alpha=\pm,\beta=e,h}
 \phi_{\alpha\beta} 
 (A_{\alpha\beta}e^{iK_{\alpha\beta}z}+B_{\alpha\beta}e^{-iK_{\alpha\beta}z}),
 \label{eq:wavefunctionweylsemimetal}
\end{align}
where
$K^2_{\pm e}=K_0^2\pm 2m[(E+\mu_\text{W})^2-\lambda^2(k_x^2+k_y^2)]^{1/2}$,
$K^2_{\pm h}=K_0^2\pm 2m[(E-\mu_\text{W})^2-\lambda^2(k_x^2+k_y^2)]^{1/2}$, and eigenvectors are
\begin{align*}
 &\phi_{+e} 
 \propto
 \begin{pmatrix}
  s_e \\
  \lambda k_-\\
  0\\
  0
 \end{pmatrix},\,
 \phi_{-e} 
 \propto
 \begin{pmatrix}
  \lambda k_+\\
  s_e \\
  0\\
  0
 \end{pmatrix},\,
 \phi_{+h} 
 \propto
 \begin{pmatrix}
  0\\
  0\\
  \lambda k_-\\
  s_h
 \end{pmatrix},\,
 \phi_{-h} 
 \propto
 \begin{pmatrix}
  0\\
  0\\
  s_h \\
  \lambda k_+
 \end{pmatrix},
\end{align*}
where $s_e=E+\mu_\text{W}+[(E+\mu_\text{W})^2-\lambda^2(k_x^2+k_y^2)]^{1/2}$, $s_h=E-\mu_\text{W}+[(E-\mu_\text{W})^2-\lambda^2(k_x^2+k_y^2)]^{1/2}$, and $k_{\pm}=k_y\pm ik_x$.

A superconductor layer lies in $z\in(-b,0)$, and the Hamiltonian is given by
\begin{align}
 H_\text{SC}(k_x,k_y,-i\partial_z)
 =
 \begin{pmatrix}
  \xi \sigma^0 & i\Delta\sigma^y \\
  -i\Delta^{\ast}\sigma^y & -\xi\sigma^0
 \end{pmatrix},
\end{align}
where $\xi=(k_x^2+k_y^2-\partial_z^2)/2m'-\mu_\text{S}$.
Hereafter, the pair potential $\Delta$ is assumed to be real without loss of generality.
The eigenfunction for an eigenenergy $E$ is given by
\begin{align}
 \psi(z)
 =&
 \sum_{\alpha=\pm,\gamma=\uparrow,\downarrow}
 \phi_{\alpha\gamma}
 (C_{\alpha\gamma}e^{iQ_{\alpha}z}+D_{\alpha\gamma}e^{-iQ_{\alpha}z}),
 \label{eq:wavefunctionsuperconductor}
\end{align}
where $Q^2_{\pm}=2m'[\mu_\text{S}\pm(E^2-|\Delta|^2)^{1/2}]-k_x^2-k_y^2$, and eigenvectors are
\begin{align*}
 &\phi_{+\uparrow}
 =
 \begin{pmatrix}
  u\\
  0\\
  0\\
  v
 \end{pmatrix},\, 
 \phi_{-\uparrow}
 =
 \begin{pmatrix}
  v\\
  0\\
  0\\
  u
 \end{pmatrix},\,
 \phi_{+\downarrow}
 =
 \begin{pmatrix}
  0\\
  u\\
  -v\\
  0
 \end{pmatrix},\, 
 \phi_{-\downarrow}
 =
 \begin{pmatrix}
  0\\
  v\\
  -u\\
  0
 \end{pmatrix}.
\end{align*}
Here, $u^2=[1+(E^2-\Delta^2)^{1/2}/E]/2$ and $v^2=[1-(E^2-\Delta^2)^{1/2}/E]/2$.

The eigenfunction in the whole unit cell $z\in(-b,a]$ is constructed by determining coefficients $A_{\alpha\beta},B_{\alpha\beta},C_{\alpha\gamma}$ and $D_{\alpha\gamma}$ so that the wave functions (\ref{eq:wavefunctionweylsemimetal}) and (\ref{eq:wavefunctionsuperconductor}) satisfy boundary conditions at $z=0$ and $z=a=-b$.
Boundary conditions for a momentum $k_z$ are given by requiring continuity of the wave function
\begin{align}
 \psi(+0)=\psi(-0),\quad
 \psi(a-0)=\psi(-b+0)e^{ik_z(a+b)},
 \label{eq:boundarycondition12}
\end{align}
and probability-current conservation
\begin{align}
 &\frac{\sigma^z\tau^z}{2m}\frac{d\psi}{dz}(+0)
 =
 \frac{\tau^z}{2m'}\frac{d\psi}{dz}(-0) + h\tau^z\psi(0),
 \label{eq:boundarycondition3}\\
 &\frac{\sigma^z\tau^z}{2m}\frac{d\psi}{dz}(a-0)
 =
 \frac{\tau^z}{2m'}\frac{d\psi}{dz}(-b+0)e^{ik_z(a+b)} + h\tau^z\psi(a).
 \label{eq:boundarycondition4}
\end{align}
The conditions (\ref{eq:boundarycondition3}) and (\ref{eq:boundarycondition4}) are derived by integrating $H(z)\psi(z)$ around the boundaries.
An eigenfunction for an eigenenergy $E$ and momenta $\bm{k}=(k_x,k_y,k_z)$ is present when a nonvanishing solution to the coefficients $A_{\alpha\beta},B_{\alpha\beta},C_{\alpha\gamma}$ and $D_{\alpha\gamma}$ is present. The boundary conditions form 16 linear equations which are described by a $16\times16$ matrix, and vanishing determinant of it is the condition of the presence of a solution.

Although actual inversion symmetry in the Weyl semimetal (\ref{eq:hamiltonianweylsemimetalnambu}) is broken by Rashba coupling, the Hamiltonian at momenta $\bm{k}$ and $-\bm{k}$ are related by $\sigma^zH_\text{WSM}(-\bm{k})\sigma^z =H_\text{WSM}(\bm{k})$.
Thus, combined with particle-hole symmetry, all eigenenergies come in positive- and negative-energy pairs at each momentum $\bm{k}$.

Notice that the thickness of the two layers $a$ and $b$ has a typical length scale, that is, the inverse of the momentum $K_0$ and $Q_0=|Q_{\pm}(E=0)|=[2m'(\mu_\text{S}^2+|\Delta|^2)^{1/2}]^{1/2}$ for the Weyl semimetal and superconductor, respectively, since, as will be explained later, the phase diagram is drawn in the space of $K_0a$ and $Q_0b$.
Therefore, the typical thickness of the superconductor layer $b\sim Q_0^{-1}$ is much shorter than the coherence length $\xi$, since $Q_0\xi\sim (|\Delta|/\mu_\text{S})^{-1}\gg 1$.
In general, the thickness needs to be much longer than the lattice constant of the stacking direction to guarantee the momentum along the $z$ direction to be a good quantum number.

In most realization of the Weyl semimetal, the number of the Weyl nodes is more than two, which is the possible minimum number for magnetic Weyl semimetals. 
The Weyl semimetal with only a single pair of the Weyl nodes described by (\ref{eq:hamiltonianweylelectron}) has been realized under an external magnetic field in EuCd$_2$As$_2$ \cite{soh19,wang19}. However, such a material could not be used as a WSM-SC multilayer unless the superconducting order of the superconductor layer persists under a strong magnetic field, like that of Nb or NbN.
On the other hand, Weyl semimetals with a single pair of Weyl nodes \textit{without} an external magnetic field, such as a TI-NI multilayer \cite{burkov11}, have not been realized experimentally.
Therefore, materials corresponding to our theory could be realized provided that (1) a Weyl semimetal with a single pair of the Weyl nodes in the absence of external magnetic field is discovered in the future, or (2) an s-wave superconductor with high upper-critical field is used to construct a superlattice with e.g. EuCd$_2$As$_2$.

\section{Electronic properties $(\mu_\text{W}=0)$}
\label{sec:electronicproperties}

In this section, we consider the case where the Fermi level of the Weyl semimetal lies exactly at the node $(\mu_\text{W}=0)$.
The condition for the presence of $A_{\alpha\beta},B_{\alpha\beta},C_{\alpha\gamma}$ and $D_{\alpha\gamma}$ can be reduced by restricting our discussion to $k_x=k_y=0$.
In this case, the first and fourth components of the spinor are decoupled from the remaining two components.
In the superconductor side, the former (latter) is relevant in considering  $\phi_{+\uparrow}$ ($\phi_{+\downarrow}$) and $\phi_{-\uparrow}$ ($\phi_{-\downarrow}$).
On the other hand, in the Weyl-semimetal side, the former (latter) is relevant in considering $\phi_{+e}$ and $\phi_{+h}$ when $E>0$ ($E<0$), and $\phi_{-e}$ and $\phi_{-h}$ when $E<0$ ($E>0$).
The condition for the presence of coefficients is explicitly shown in Appendix \ref{sec:eigenvaluequation}.

\subsection{Phase diagram without potential barriers}

First of all, consider the case when the potential barrier is absent.
Provided that the energy dispersion is linear around nodes (which will be discussed later), the phase diagram is determined by the presence or the absence of and also the number of nodes, which are the $E=0$ solutions of (\ref{eq:eigenvalueequation}).
When $E=0$, the momentum in the $z$ direction of the Weyl-semimetal layer is $K_0$, and let the absolute value of that of the superconductor layer be denoted by $Q_0\equiv|Q_{\pm}(E=0)|=[2m'(\mu_\text{S}^2+|\Delta|^2)^{1/2}]^{1/2}$.
Since the momentum $Q_0$ is close to the Fermi momentum $(2m'\mu_\text{S})^{1/2}$ in the absence of the superconducting order ($\Delta=0$), in the following, we refer to $Q_0$ as the Fermi momentum of the superconductor for ease of description, since the genuine Fermi momentum does not appear in the following argument. 
The phase diagram is spanned by $K_0a$ and $Q_{0}b$.
One of the parameters $K_0a$ can be tuned by the variation of the thickness $a$ or the exchange field (thus, the position of the nodes $\pm K_0$) of the Weyl-semimetal layers, and the other parameter $Q_0 b$ by the variation of the thickness $b$ or the chemical potential of the superconductor layers.
The zero-energy states are always doubly degenerate, since two decoupled equations ($r=\pm 1$ in (\ref{eq:eigenvalueequation})) are identical.
The equation (\ref{eq:eigenvalueequation}) is dependent on a single quantity $\alpha^{-2}+\alpha^2$, where $\alpha=(K_0/m)/(Q_0/m')$ is the ratio of the Fermi velocity of the Weyl semimetal and (nearly) that of the superconductor.
The phase diagram is unchanged when replacing $\alpha$ by its inverse.

\begin{figure}[t]
 \centering
   \includegraphics[width=42mm]{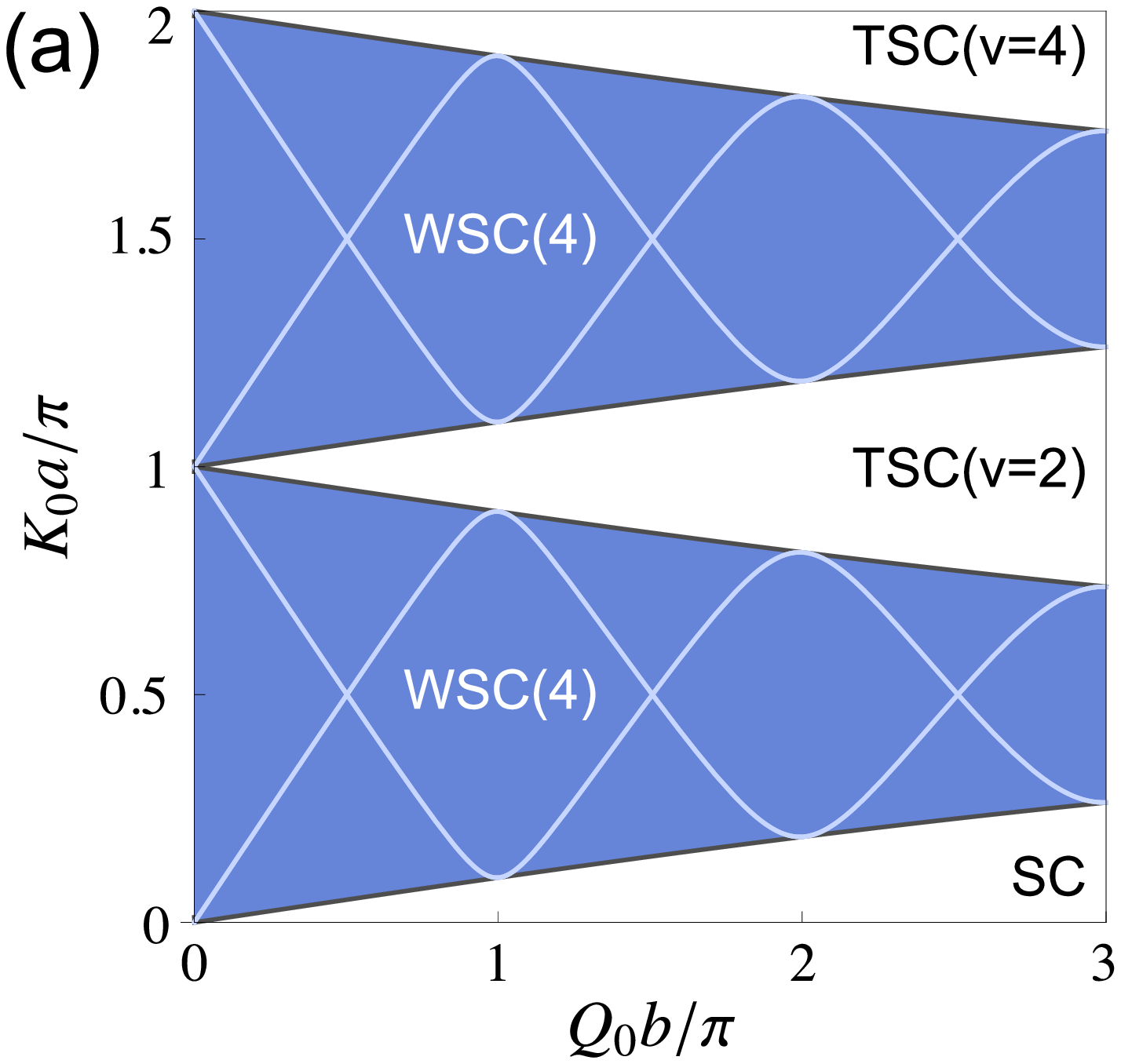}
   \includegraphics[width=42mm]{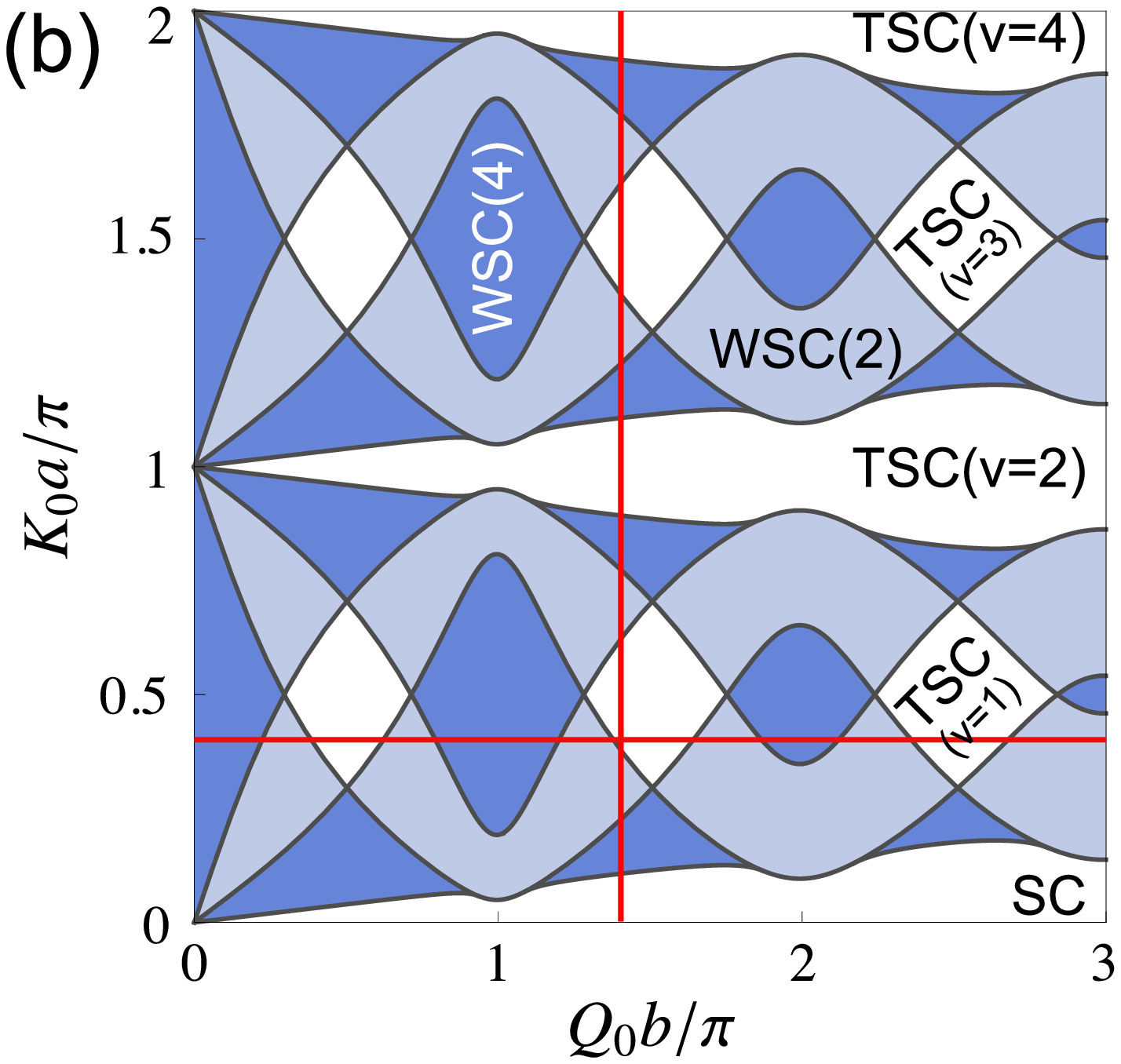}\\
   \includegraphics[width=42mm]{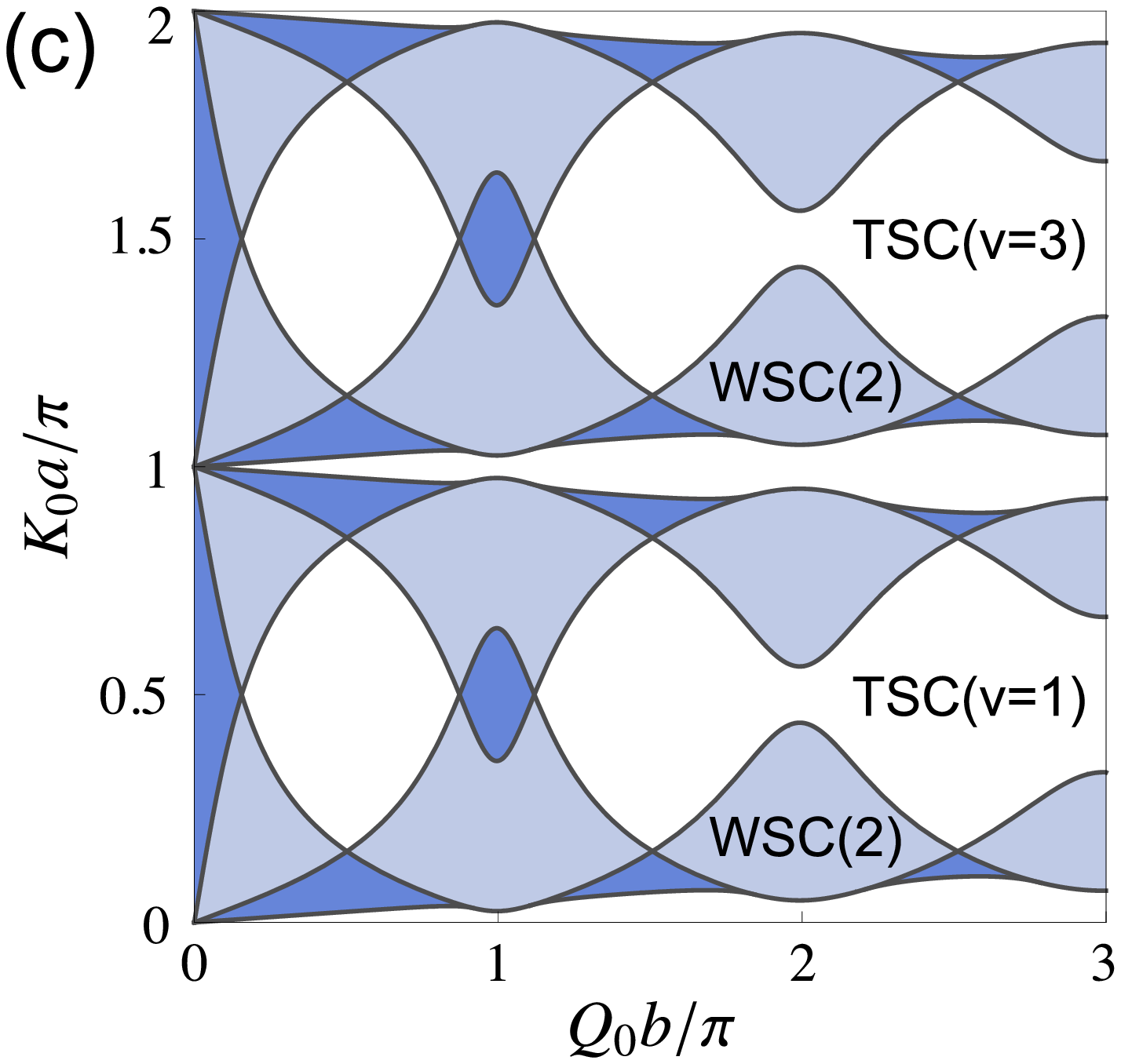}
   \includegraphics[width=42mm]{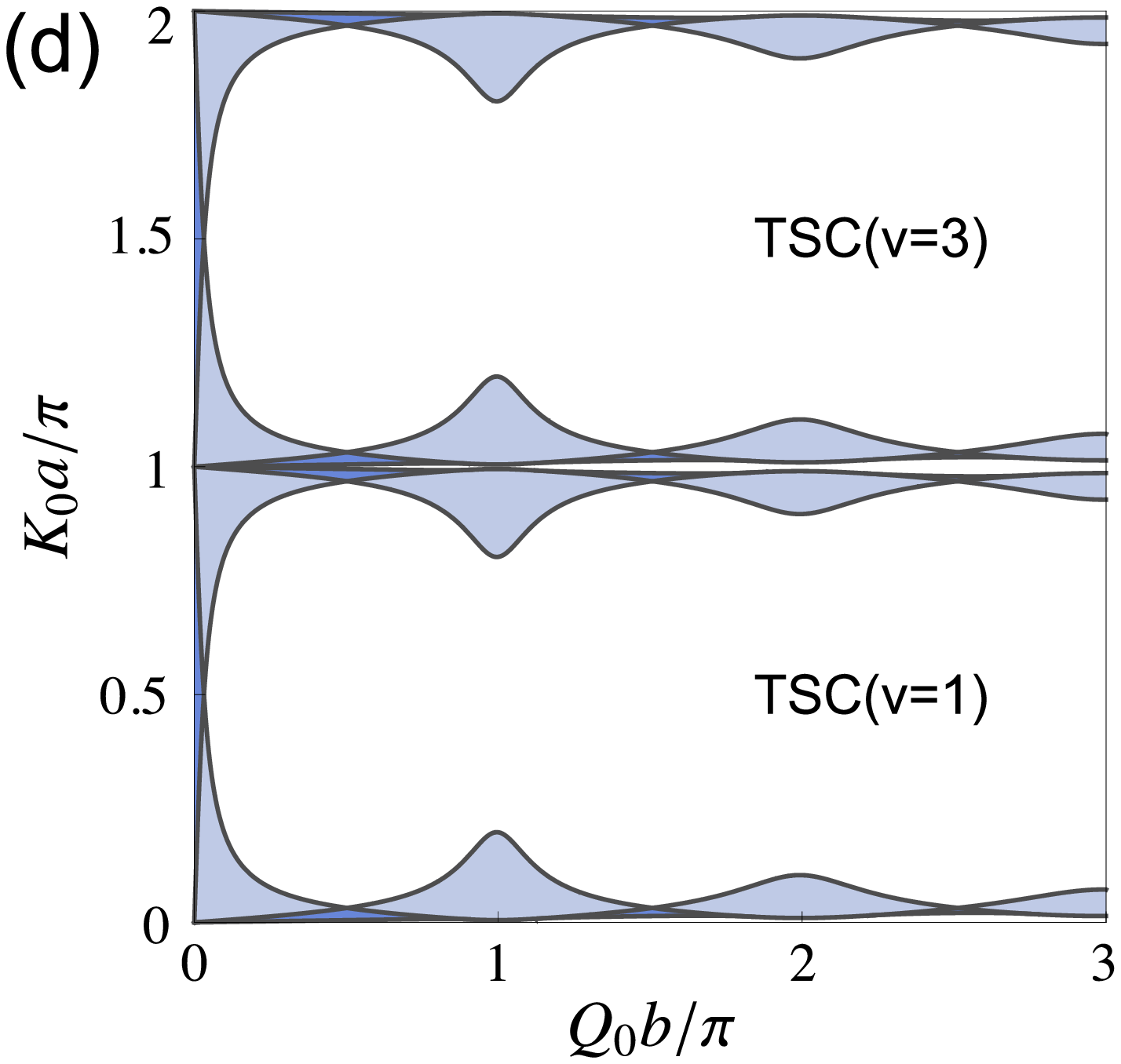}
 \caption{(Color online) The phase diagram in $Q_0b/\pi$-$K_0a/\pi$ space is shown when the ratio of the Fermi velocity $\alpha=(K_0/m)/(Q_0/m')$ is (a) $1$, (b) $1/2$, (c) $1/4$, and (d) $1/20$.
 Here, the magnitude of the pair potential is set to be $\tan^{-1}\Delta/\mu_\text{S}=0.2$.
 Dark- and light-blue regions correspond to Weyl-superconductor phases with 4 nodes [WSC(4)] and 2 nodes [WSC(2)], respectively, and white regions to fully-gapped trivial- and topological-superconductor phases (SC and TSC).
 \label{fig:phasediagram}}
\end{figure}

First, let us focus on the special case $\alpha=1$, that is, the case when two Fermi velocities coincide [Fig.~\ref{fig:phasediagram}(a)].
The Weyl nodes are gapped out by the proximity effect from even infinitesimally thin superconductor layers ($b\ll a$) when the position of the two Weyl nodes $\bm{k}=(0,0,\pm K_0)$ are identical in the Brillouin zone $k_z\in[-\pi/a,\pi/a]$.
This fully-gapped superconductor is a three-dimensional extension of a time-reversal-symmetry-broken topological superconductor, the section of which along $k_z$ is characterized by the Chern number. 
Otherwise, the two Weyl nodes are split into four nodes of the Majorana fermion to become a Weyl superconductor.
The phase boundaries between Weyl- and topological-superconductor phases are given by $\cos^2K_0a = \cosh^2Q_0b\sin\phi$, where $\tan 2\phi=\Delta/\mu_\text{S}$, and thus topological-superconductor phases become thinner (Weyl-superconductor phases become broader) as the magnitude of the pair potential $|\Delta|$ becomes smaller.

The Chern number of topological-superconductor phases can be determined by referring to the Weyl-semimetal phase. 
Along the vertical axis ($b=0$) of Fig.~\ref{fig:phasediagram}(a), the Weyl semimetal has the Hall conductivity $\sigma_{xy}=(e^2/h)K_0/\pi$.
In other words, the integral of the Chern number over $k_z$ in the Weyl semimetal is $2K_0$, which is the interval of the nodes.
When a topological-superconductor phase with the Chern number $\nu$ touches the Weyl-semimetal phase, the integral of the Chern number over $k_z$, that is, $2\pi\nu/a$ has to agree with twice that of the Weyl semimetal, since the Chern number of the energy bands of the Bogoliubov-de Gennes Hamiltonian for superconductors is doubled due to the Nambu space of the electric and hole energy bands. Thus, we obtain
\begin{align}
 \nu=2K_0a/\pi.
 \label{eq:chernnumber_phasediagram}
\end{align}

As $\alpha$ varies from unity, that is, when the Fermi velocities of the Weyl semimetal and the superconductor mismatch, Majorana nodes hybridize to have a finite energy gap when two of them meet in the momentum space, which occurs on the (light blue) solid lines inside the Weyl-superconductor phases in Fig.~\ref{fig:phasediagram}(a).
Depending on the magnitude of the energy gap, a Weyl-superconductor phase with 4 Majorana nodes turns into a Weyl-superconductor phase with 2 Majorana nodes or a fully gapped phase [Fig.~\ref{fig:phasediagram}(b), (c)].

\begin{figure}
 \centering
 \begin{tabular}{cc}
  \includegraphics[width=42mm]{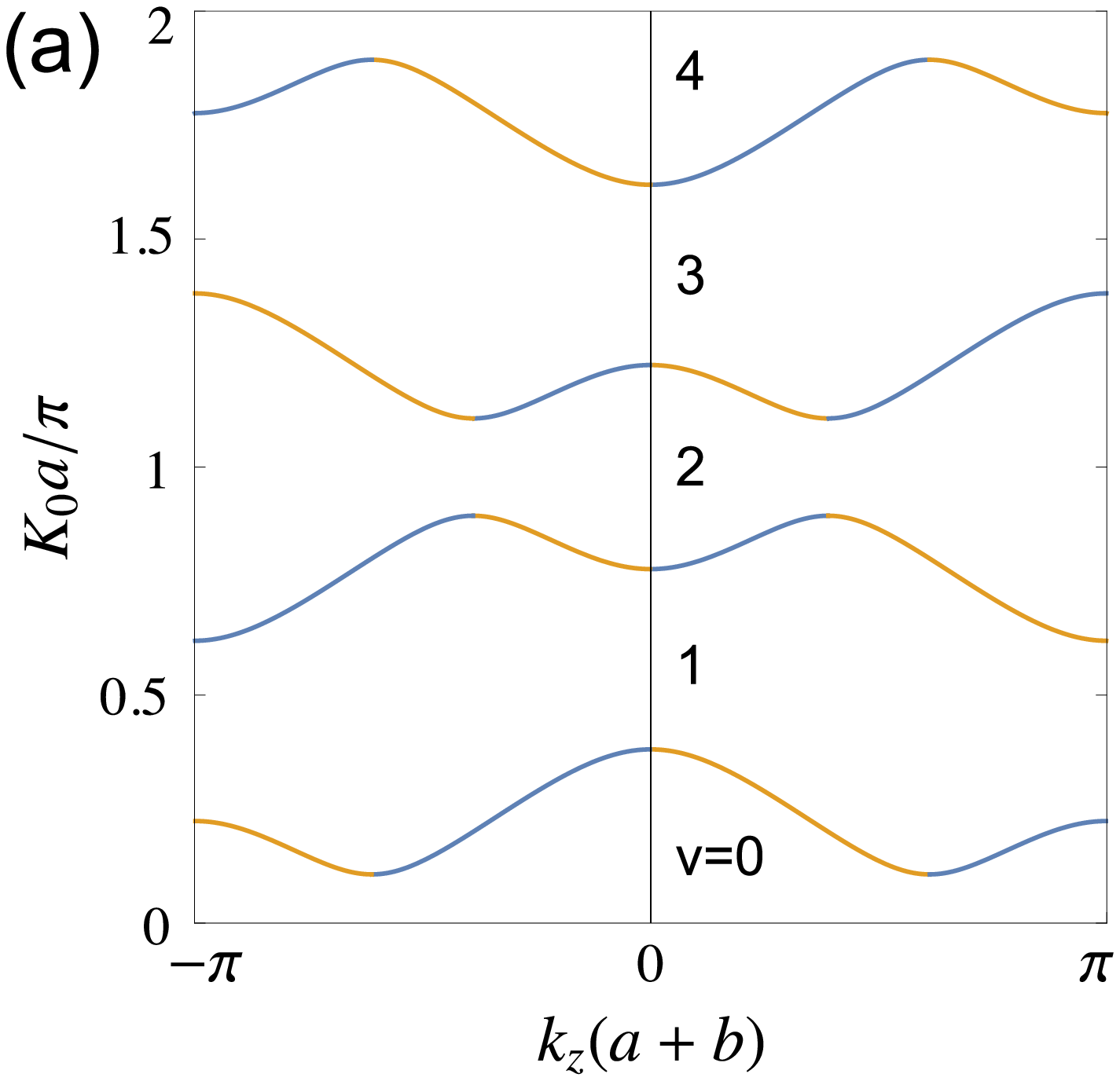}
  \includegraphics[width=42mm]{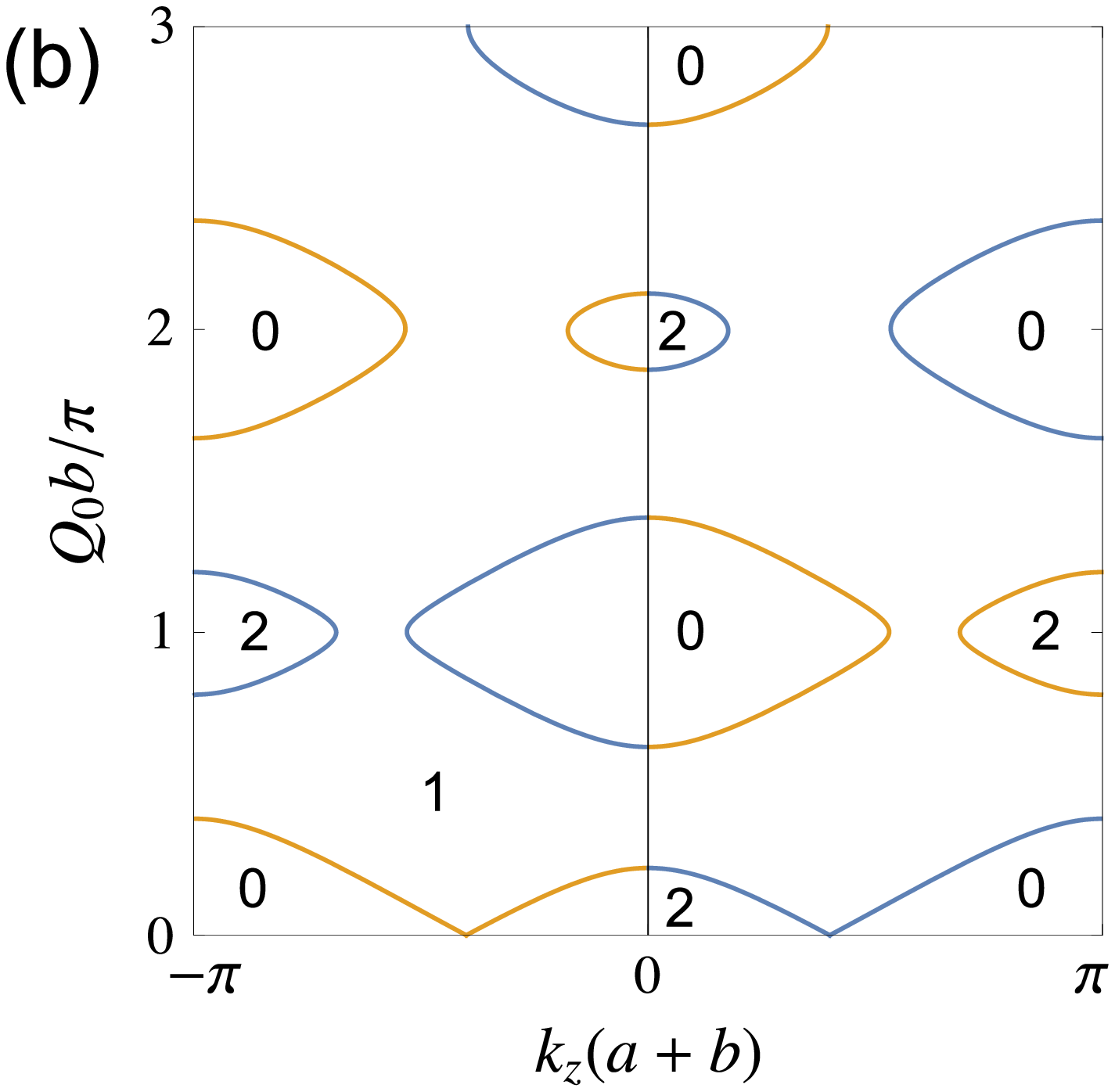}\\
  \includegraphics[width=84mm]{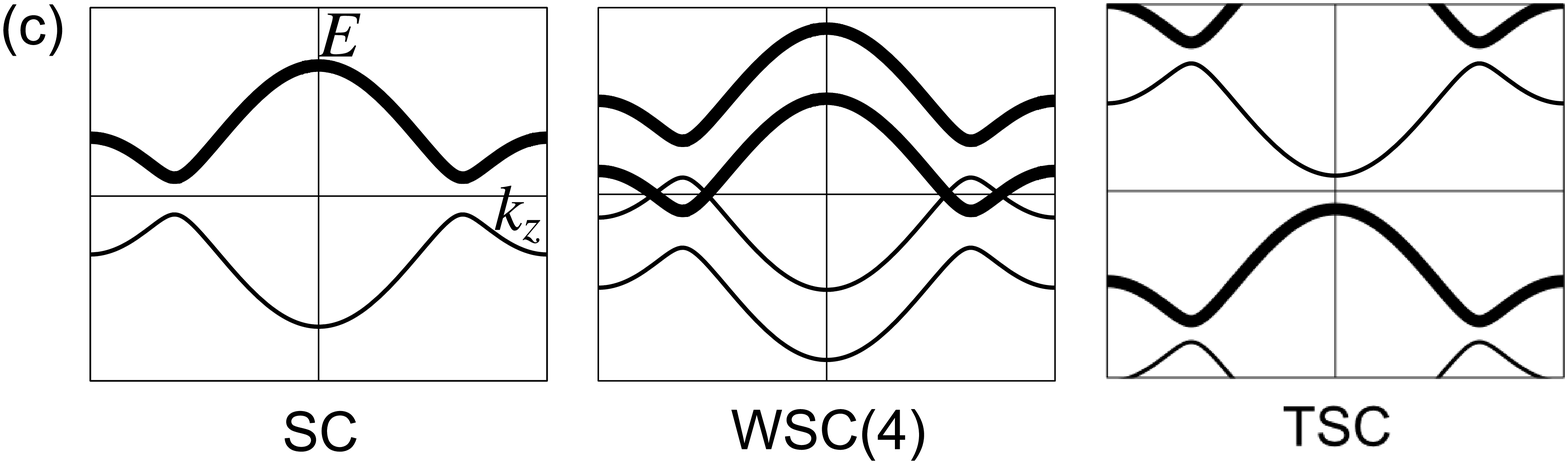}\\
  \includegraphics[width=84mm]{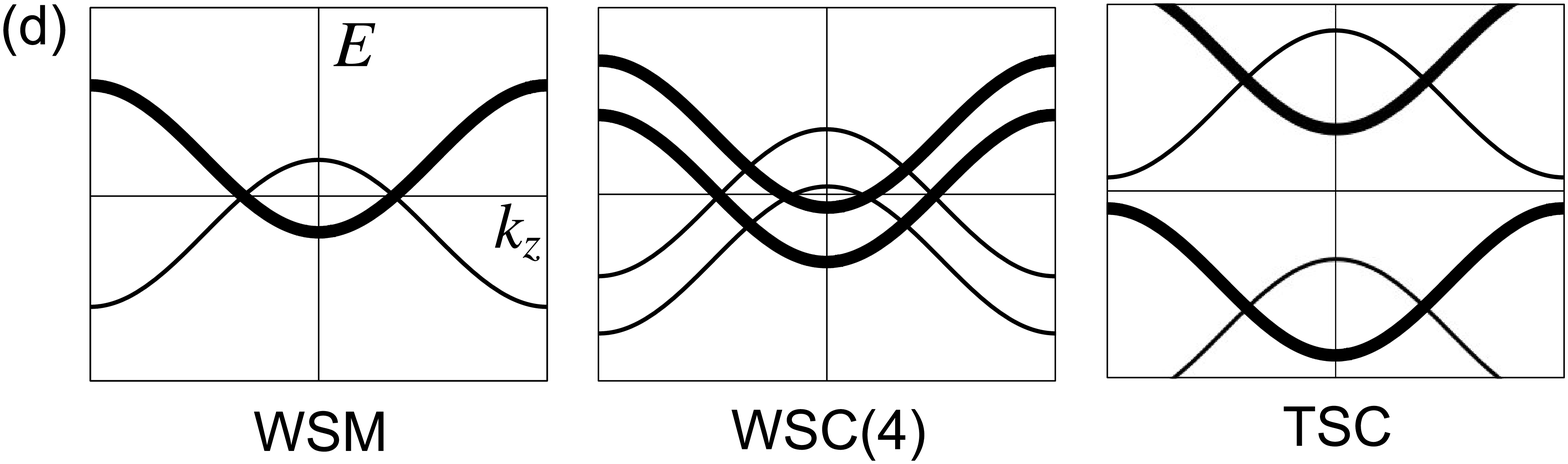}
 \end{tabular}
 \caption{(Color online) The position of Weyl nodes when $\alpha=(K_0/m)/(Q_0/m')=1/2$ is shown as a function of (a) $K_0a$ at $Q_0b/\pi=1.4$ [the vertical red line in Fig.~\ref{fig:phasediagram}(b)], and of (b) $Q_0 b$ at $K_0 a/\pi=0.4$ [the horizontal red line in Fig.~\ref{fig:phasediagram}(b)]. In (a) and (b), the number shown inside regions separated by node-position lines represents the Chern number of the electronic states in the $k_x$-$k_y$ space, and the color of lines represents the monopole charge (positive for blue lines and negative for yellow lines).
 Schematic pictures of the change of energy bands along $k_z$ are shown for the evolution (c) from a trivial superconductor (SC) at $K_0a/\pi=0$ to a Weyl superconductor with 4 nodes [WSC(4)], a Weyl superconductor with 2 nodes [WSM(2)] (not shown), and to a topological superconductor (TSC) at $K_0a/\pi=0.5$, and (d) from a Weyl semimetal (WSM) at $Q_0 b/\pi=0$  to WSC(4), WSC(2), and TSC at $Q_0 b/\pi=0.5$.
 \label{fig:positionofweylnodes}}
\end{figure}
The position of the Majorana nodes is shown in Fig.~\ref{fig:positionofweylnodes}(a) as a function of $K_0a$ at $Q_0b/\pi=1.4$ [the vertical red line in Fig.~\ref{fig:phasediagram}(b)], and in Fig.~\ref{fig:positionofweylnodes}(b) as a function of $Q_0b$ at $K_0a/\pi=0.4$ [the horizontal red line in Fig.~\ref{fig:phasediagram}(b)].
While, for the Bogoliubov-de Gennes Hamiltonian of Weyl semimetals, the Chern number of the section along $k_z$ changes by 2 in a discrete manner as $k_z$ moves across a Weyl node, the Chern number changes by unity in the case of Weyl superconductors, since the Majorana fermion is a half of the electron.
As $K_0 a/\pi$ varies from 0 to 0.5, two pairs of Majorana nodes appear, sweep once the Brillouin zone, and disappear in pairs, the process of which is schematically shown in Fig.~\ref{fig:positionofweylnodes}(c). Through this process, a trivial-superconductor phase continuously turns into a topological-superconductor phase with the Chern number 1. Therefore, the fully gapped phases that emerge within Weyl-superconductor phases in Fig.~\ref{fig:phasediagram}(b) are topological-superconductor phases with odd Chern numbers, which cannot be continuously connected to non-superconducting phases without closing the energy gap.

On the other hand, as $Q_0 b/\pi$ evolves from 0 to 0.5, a pair of Weyl nodes are split into four Majorana nodes, and then annihilate in pairs [Fig.~\ref{fig:positionofweylnodes}(b)]. This process is schematically shown in Fig.~\ref{fig:positionofweylnodes}(d), and shares the same scenario as the intrinsic superconductivity in Weyl semimetals shown in (\ref{eq:weylsemimetal_sc})-(\ref{eq:modelwsmsc_eigenenergy}). By further increasing $Q_0 b/\pi$, the Chern number remains at most 2, since a pair of created Majorana nodes annihilate on their own, and do not sweep the Brillouin zone.

As $\alpha$ becomes far larger (or equivalently far smaller) than unity, topological-superconductor phases with even Chern numbers and Weyl-superconductor phases with 4 Majorana nodes shrink, and the phase diagram is almost dominated by topological-superconductor phases with odd Chern numbers [Fig.~\ref{fig:phasediagram}(d)].
This indicates that mismatch of the Fermi velocity favors topological-superconductor phases with odd Chern numbers.

\subsection{Energy dispersion near the Majorana nodes}
\label{sec:degenerateperturbationtheory}

So far, electronic properties are discussed relying on zero-energy states.
In order to identify Weyl-superconductor phases, we need to confirm that the low-energy properties are described by the Weyl Hamiltonian.

We estimate the energy dispersion away from $k_x=k_y=0$ by the degenerate perturbation theory.
Let the position of a node be given by $\bm{k}^{(0)}=(0,0,k_z^{0})$, and its deviation be denoted by $\bm{p}=\bm{k}-\bm{k}^{(0)}$.
Unperturbed wave functions are those on the node.
Up to linear order in $\bm{p}$, the effective Hamiltonian has the form of
\begin{align}
 H_\text{eff}
 =
 \begin{pmatrix}
  v_1p_z & v_2(p_y+ip_x)\\
  v_2(p_y-ip_x) & -v_1p_z
 \end{pmatrix},
 \label{eq:effectivehamiltonian_mainbody}
\end{align}
the detailed expression of which is shown in Appendix \ref{sec:degenerateperturbation}.
The expression of the effective Hamiltonian agrees with the fact that the Chern number changes by unity when $k_z$ moves across the Majorana node since the diagonal elements work as a mass in $p_x$-$p_y$ plane which changes its sign at the node, provided that $v_2$ is nonvanishing. The Chern number is constant in the interval between two nodes [Fig.~\ref{fig:positionofweylnodes}(a), (b)].
The monopole charge is determined by the sign of the velocity $v_1$.
As can be seen in Fig.~\ref{fig:effectivehamiltoniancoefficients}, pair-created and pair-annihilated Majorana nodes have the opposite signs of the velocity $v_1$, which indicates that a pair of Majorana nodes appear and disappear with the opposite monopole charge.
It is also numerically checked that the velocity $v_1$ agrees with the exact dispersion along $k_z$ derived from (\ref{eq:eigenvalueequation}) near nodes.

\subsection{Thermal Hall conductivity}

One of the topological nature in Weyl semimetals appears in the Hall conductivity.
In a superconductor, characteristic Hall effect is realized in thermal transport, that is, the thermal Hall (Righi-Leduc) effect \cite{meng12}.
In Weyl superconductors, the thermal Hall conductivity $\kappa_{xy}$ is given by the integral of the Chern number over $k_z$ multiplied by $\kappa_{xy}^0/2\pi$, since the thermal Hall conductivity of a two-dimensional topological superconductor with the Chern number $\nu$ is quantized as $\kappa_{xy}=\nu\kappa_{xy}^0$ \cite{read00,nomura12,sumiyoshi13}. Here, $\kappa_{xy}^0=\pi^2k_\text{B}^2T/6h$ is the quantum of the thermal Hall conductivity that is realized in a two-dimensional topological superconductor with the Chern number $\nu=1$ (or, equivalently, with the chiral central charge $c=1/2$), and is half the thermal Hall conductivity $\kappa_{xy}^{0,\text{QH}}=\pi^2k_\text{B}^2T/3h$ of a two-dimensional quantum Hall system with the Chern number $\nu=1$ from the Wiedemann-Franz law \cite{smrcka77}.

The thermal Hall conductivity is deduced from the interval of Majorana nodes in Weyl-superconductor phases and from the Chern number in topological-superconductor phases.
\begin{figure}
 \centering
 \includegraphics[width=60mm]{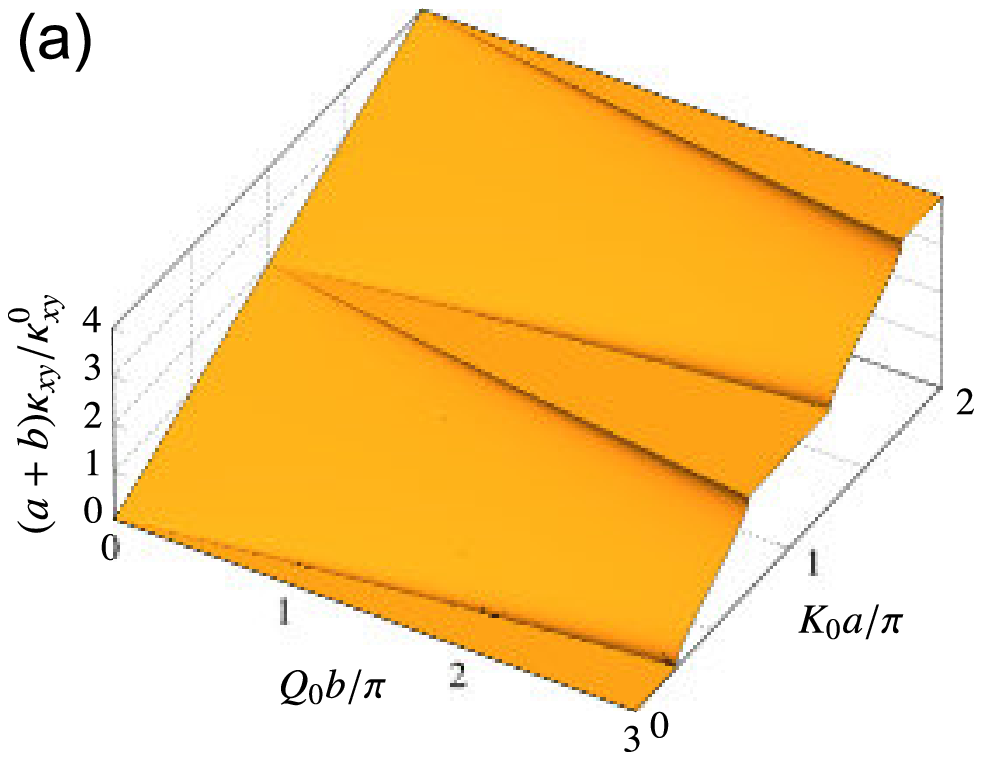}
 \includegraphics[width=60mm]{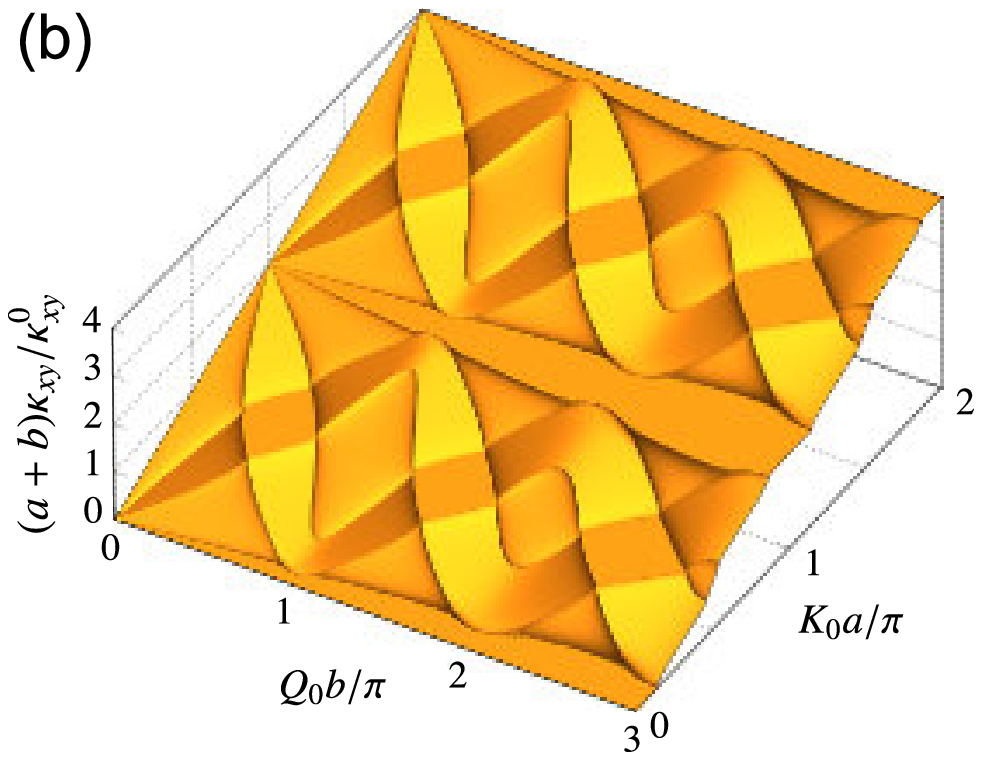}
 \includegraphics[width=60mm]{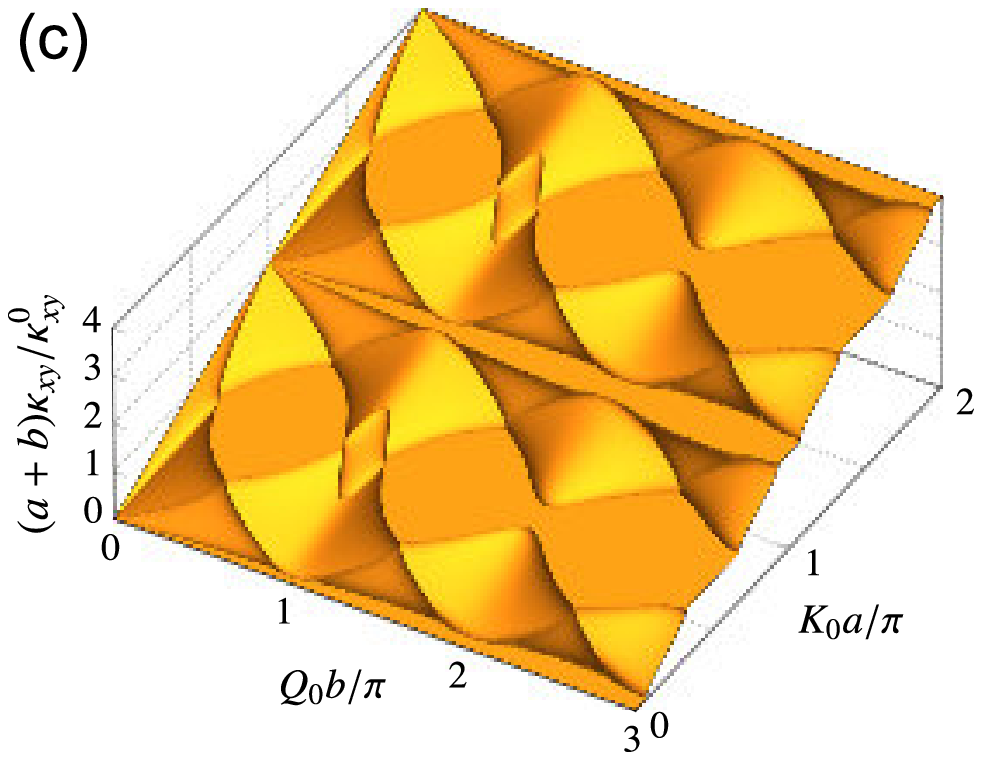}
 \includegraphics[width=60mm]{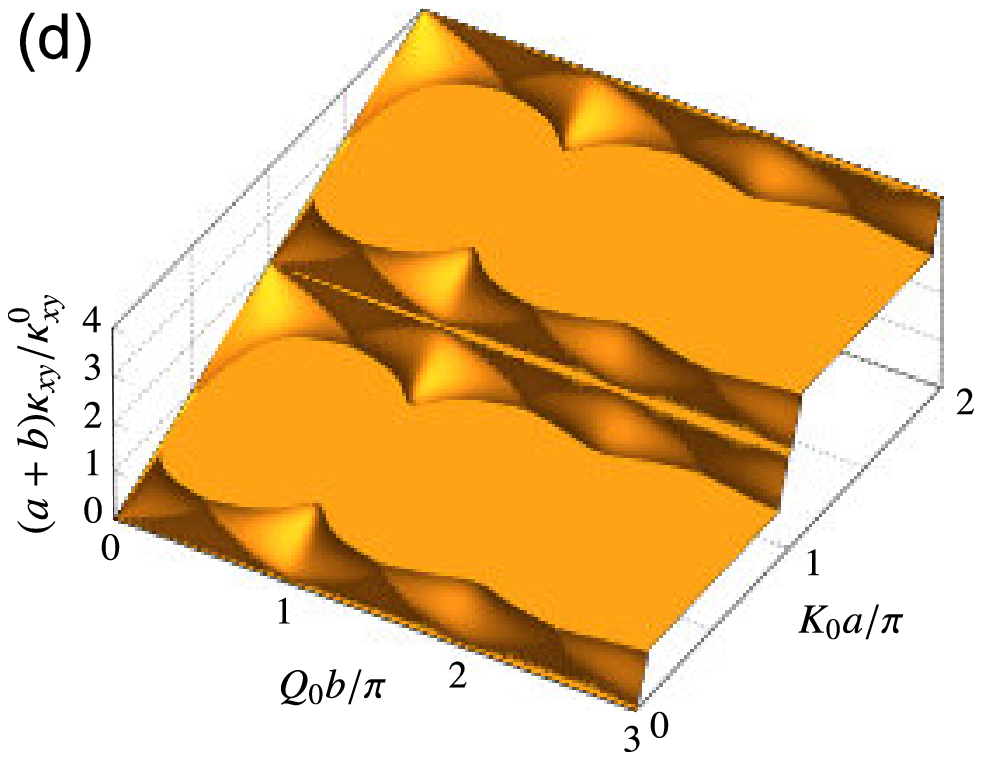}
 \caption{The thermal Hall conductivity $\kappa_{xy}$ multiplied by the unit-cell size $a+b$ is shown in units of the two-dimensional thermal Hall conductivity quantum $\kappa_{xy}^0=\pi^2 k_\text{B}^2T/6h$. Each figure corresponds to the ratio of the Fermi velocity $\alpha=(K_0/2m)(Q_0/2m')$ being (a) $1$, (b) $1/2$, (c) $1/4$, and (d) $1/20$. The thermal Hall conductivity is estimated by the interval of nodes. 
 \label{fig:hallconductivity}}
\end{figure}
The thermal Hall conductivity is shown in Fig.~\ref{fig:hallconductivity} for the same parameter used in Fig.~\ref{fig:phasediagram}(a)-(d).
In topological-superconductor phases, the thermal Hall conductivity multiplied by the unit-cell size $a+b$ is quantized by integers, or equivalently, by integers and half-odd integers by measuring in units of $\kappa_{xy}^{0,\text{QH}}$.
On phase boundaries, the thermal Hall conductivity shows a kink, since Majorana nodes move faster in the momentum space as a function of $K_0a$ or $Q_0b$ when close to phase boundaries, which can be seen in Fig.~\ref{fig:positionofweylnodes}(a) and (b).

\subsection{Phase diagram with potential barriers}

When potential barriers are present at the interface of Weyl-semimetal and superconductor layers, an electron in a Weyl-semimetal layer is more likely to be reflected at the interface, which results in an energy gap when two Majorana nodes with the opposite monopole charge are closely positioned. Therefore, the potential barrier have a similar effect as the mismatch of the Fermi velocity.

\begin{figure}
 \centering
 \includegraphics[width=42mm]{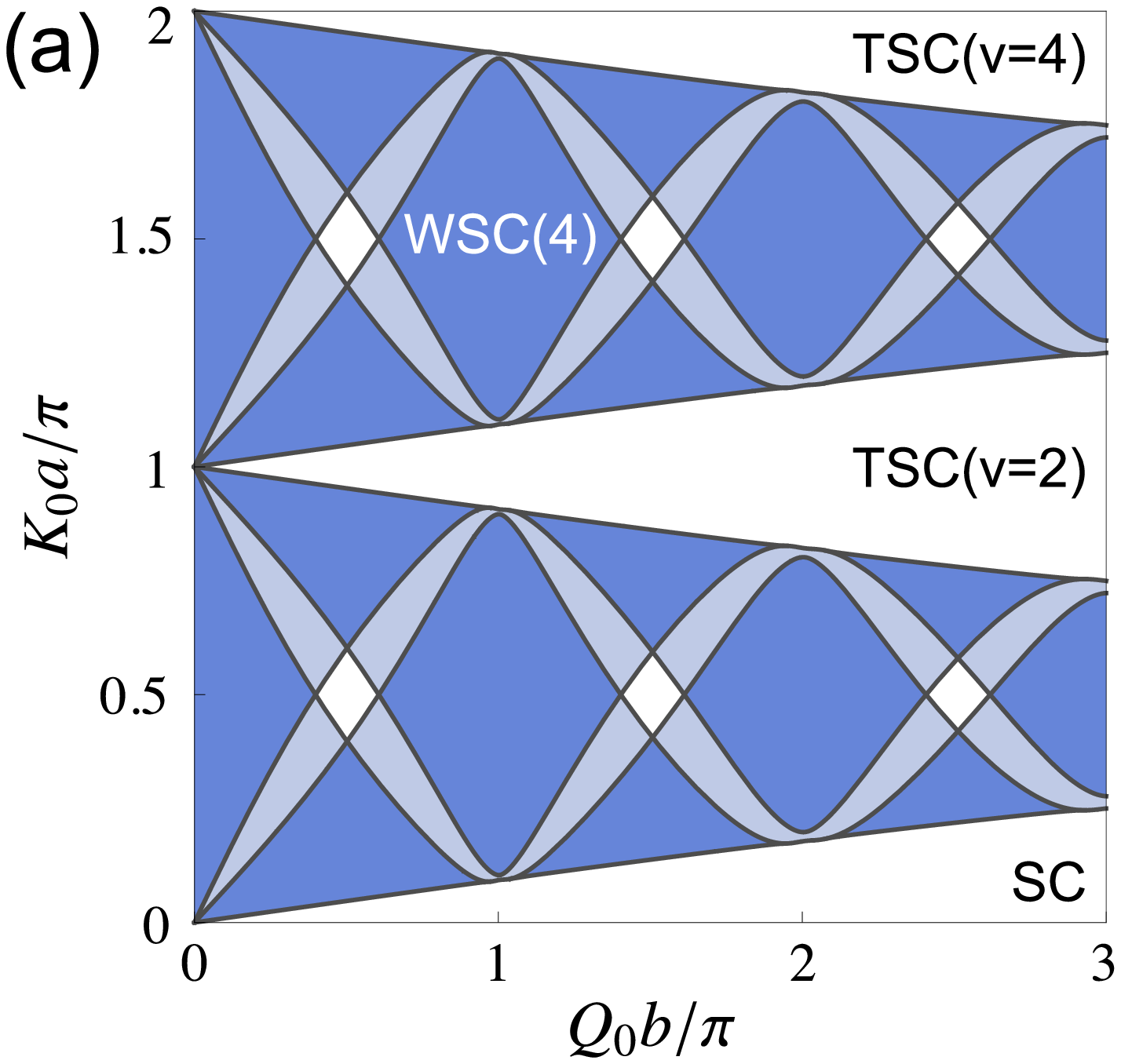}
 \includegraphics[width=42mm]{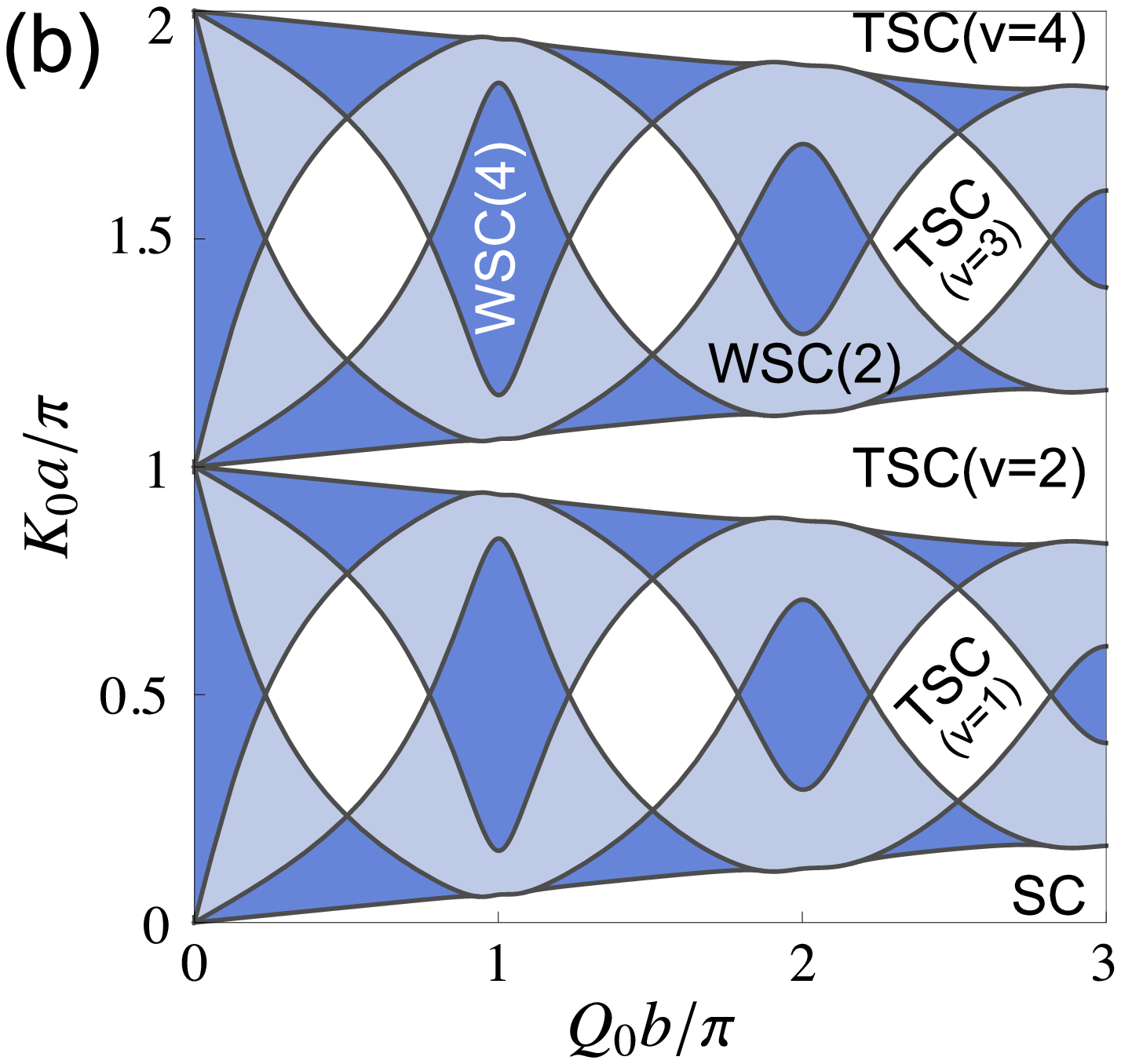}\\
 \includegraphics[width=42mm]{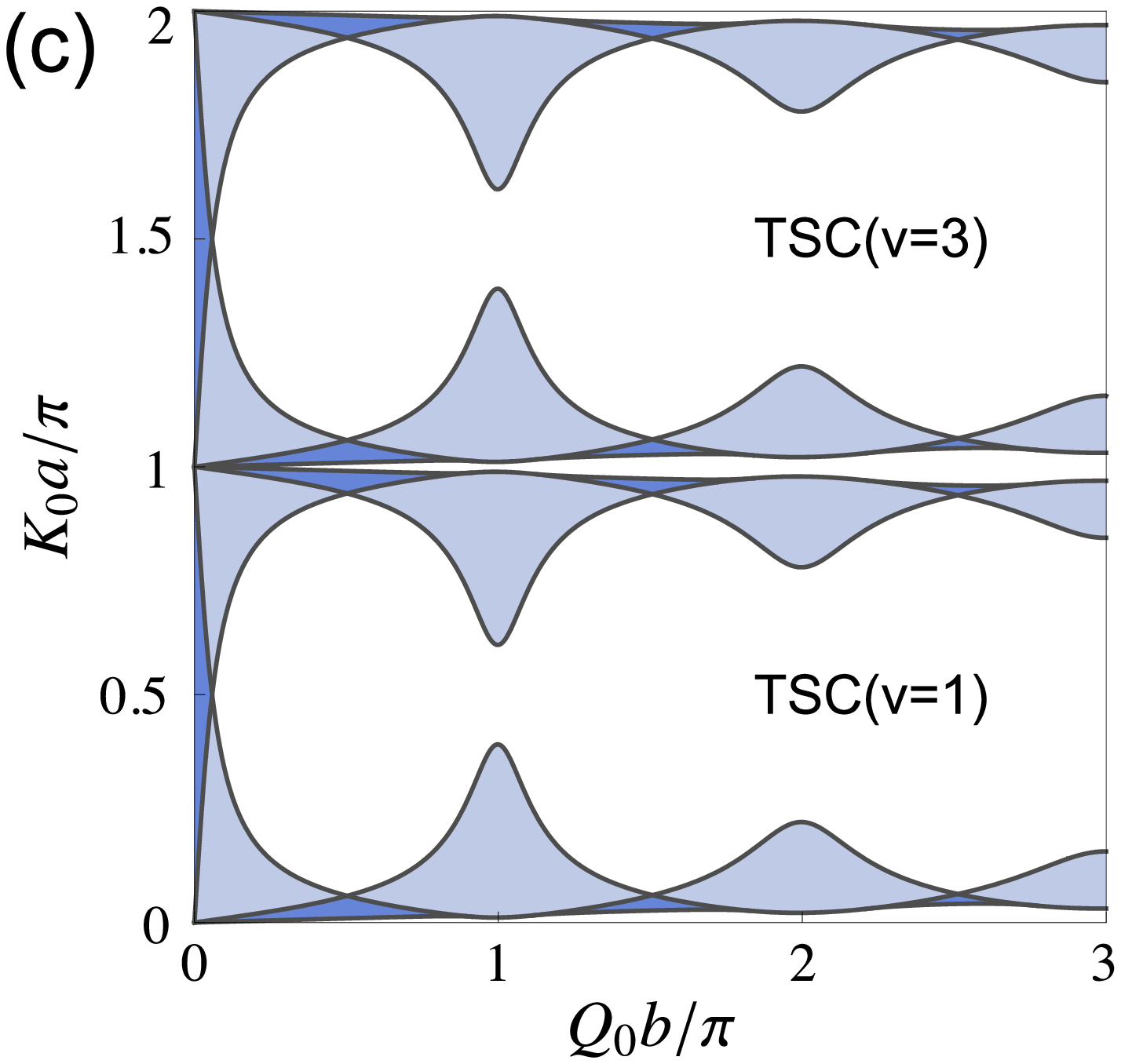}
 \caption{The phase diagram in the presence of the potential barrier is shown for the barrier height $h/(Q_0/2m')$ being (a) $1/3$, (b) 1, and (c) 3. 
 The other parameters are the same as those used in Fig.~\ref{fig:phasediagram}(a).
 \label{fig:phasediagrambarrier}}
\end{figure}

The phase diagram in the presence of the potential barrier is determined also from the zero-energy solution of (\ref{eq:eigenvalueequation}), and is shown in Fig.~\ref{fig:phasediagrambarrier}, where the corresponding phase diagram without the barrier is Fig.~\ref{fig:phasediagram}(a). The barrier height is $h/(Q_0/2m')=1/3,1,3$, respectively.
As in the case of the large Fermi-velocity mismatch, a large barrier height favors topological-superconductor phases with odd Chern numbers.

\section{Electronic properties $(\mu_\text{W}\neq 0)$}
\label{sec:awayfromthenode}
In this section, we examine the case when the Fermi level of the Weyl semimetal is away from the node ($\mu_\text{W}\neq 0$).
The phase diagram is determined from the $E=0$ solutions at $k_x=k_y=0$ as has been done in the previous section. However, the same argument as in Sec.~\ref{sec:degenerateperturbationtheory} indicates that the low-energy behavior is described by the Weyl Hamiltonian unless some components of the effective Hamiltonian (\ref{eq:effectivehamiltonian_mainbody}) vanish accidentally.

\begin{figure}
 \centering
   \includegraphics[width=42mm]{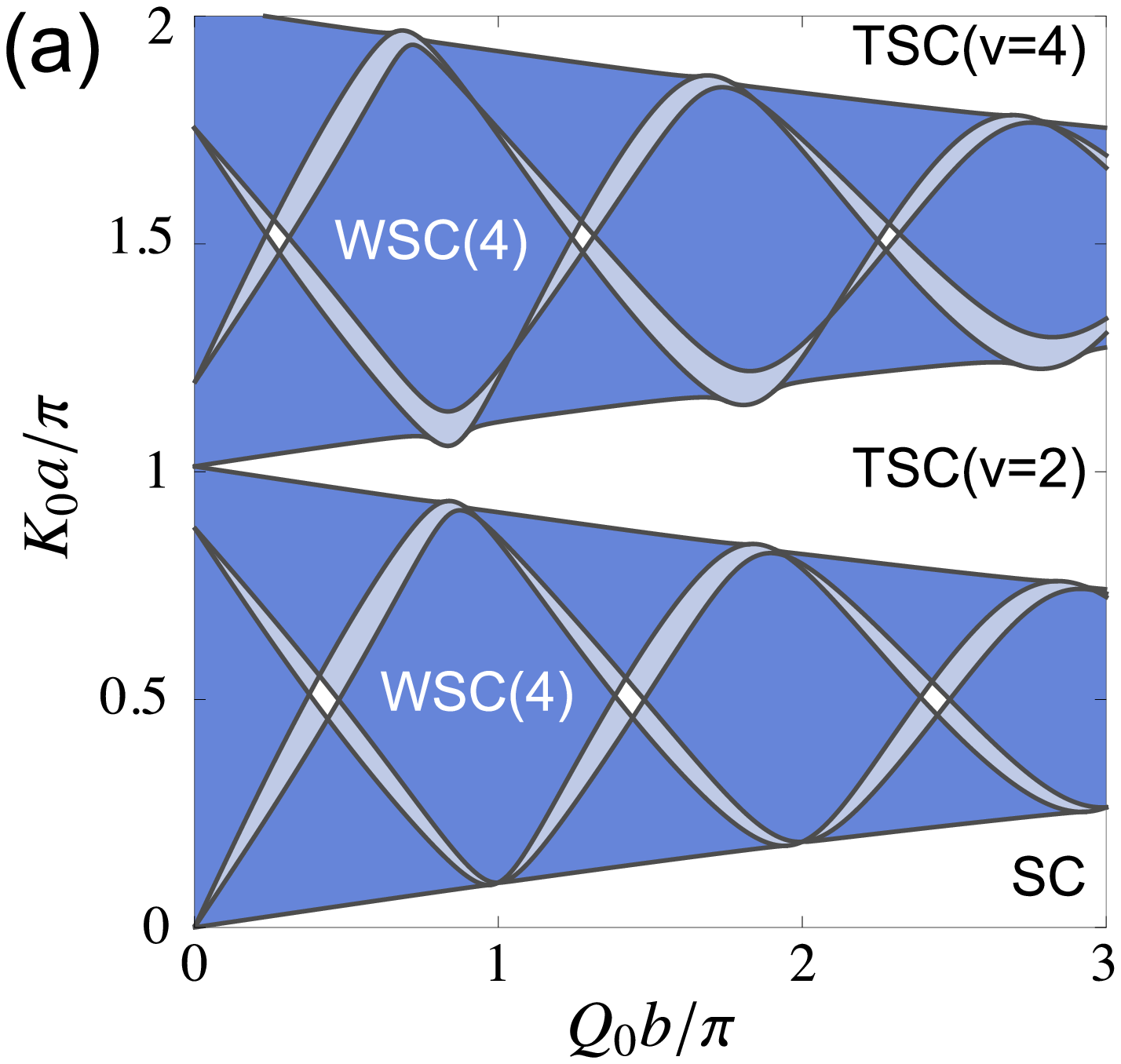}
   \includegraphics[width=42mm]{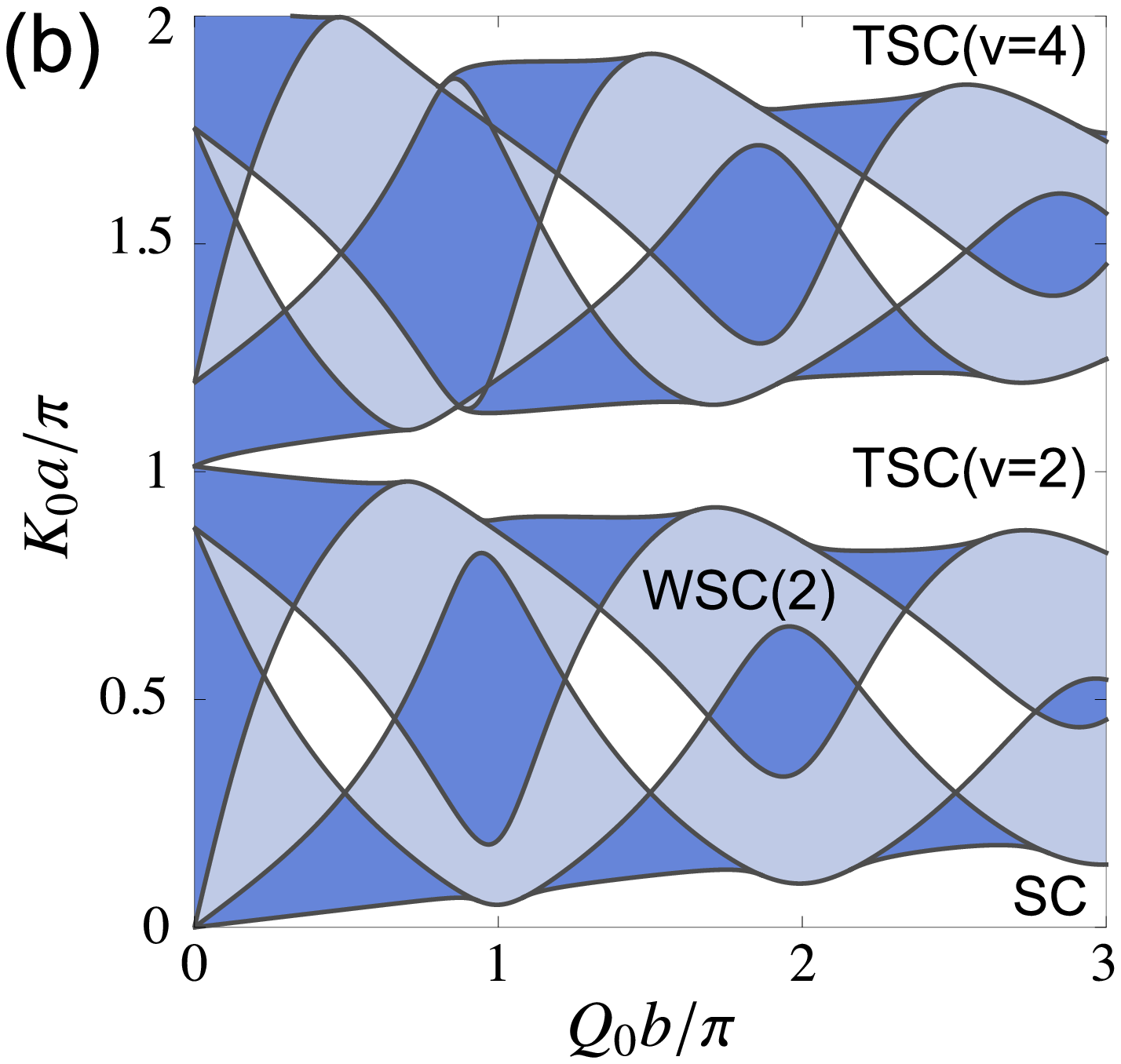}\\
   \includegraphics[width=42mm]{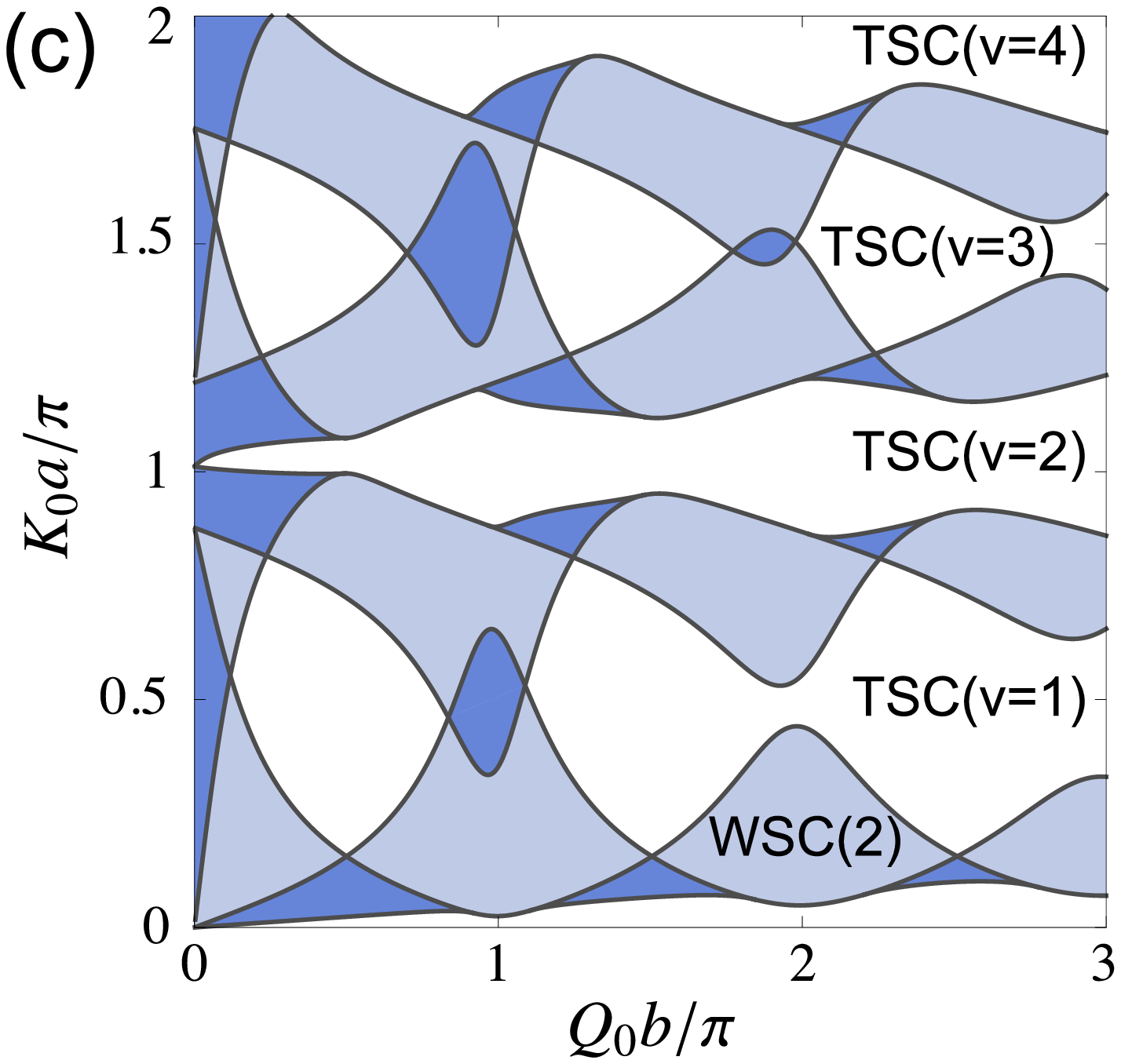}
   \includegraphics[width=42mm]{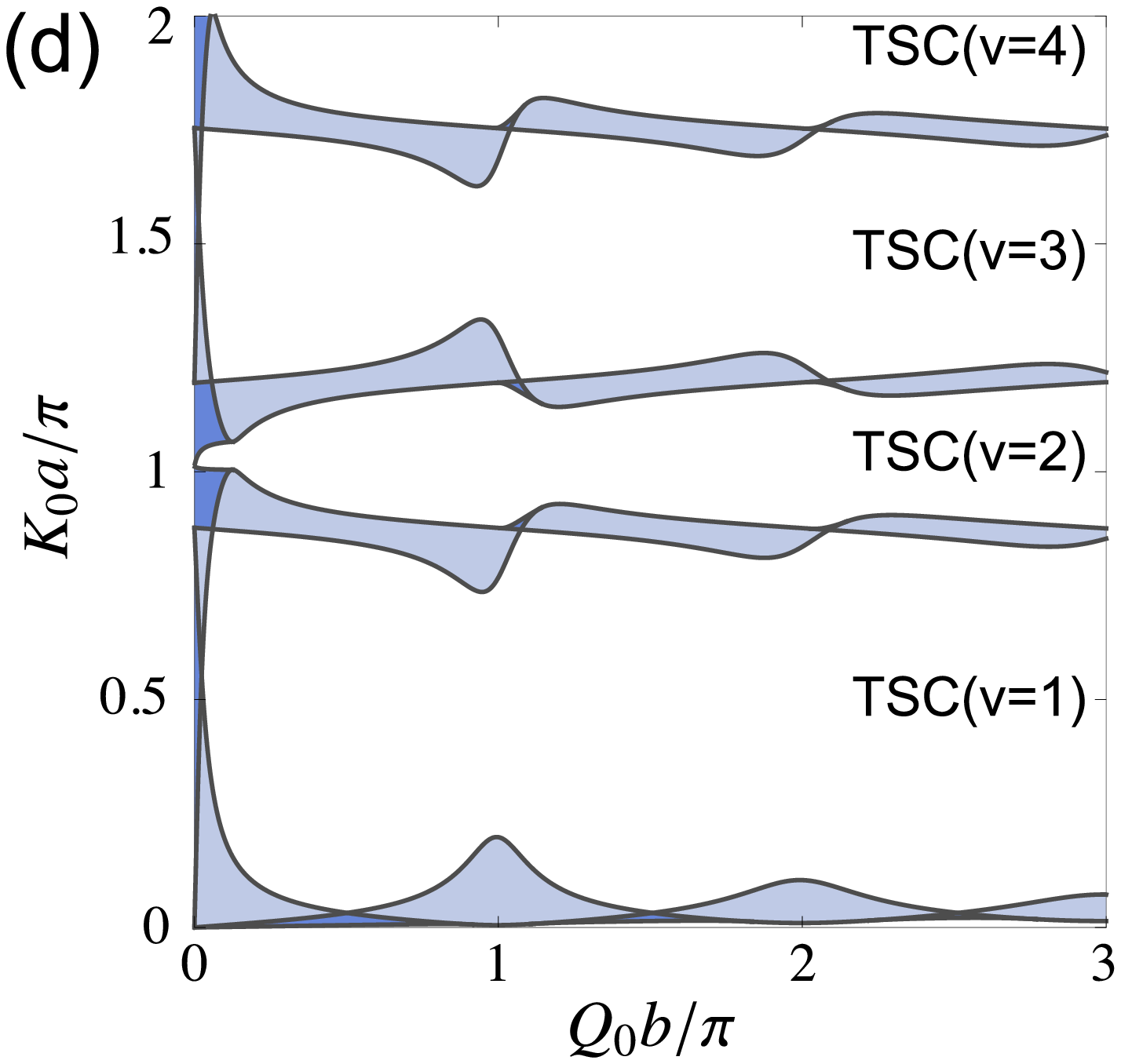}
 \caption{(Color online) The phase diagram when the Fermi level is $\mu_\text{W}=0.3K_0^2/2m$ in $Q_0b/\pi$-$K_0a/\pi$ space is shown for the ratio of the Fermi velocity $\alpha=(K_0/m)/(Q_0/m')=$  (a) $1$, (b) $1/2$, (c) $1/4$, and (d) $1/20$.
 The other parameters are the same as those in Fig.~\ref{fig:phasediagram}.
 \label{fig:phasediagram_fermilevel03}}
\end{figure}
The phase diagrams when the Fermi level of the Weyl semimetal is $\mu_\text{W}=0.3 K_0^2/2m$ and without the potential barrier ($h=0$) are shown in Fig.~\ref{fig:phasediagram_fermilevel03} for the same parameters as used in Fig.~\ref{fig:phasediagram}.
When the Fermi level is away from the node, the periodic structure of the phase diagram as a function of the Weyl-semimetal-layer thickness $K_0a$ cannot be seen, since the Fermi momenta of the spin-up and spin-down bands [$(K_0^2\pm 2m\mu_\text{W})^{1/2}$] are different.
Even in the presence of finite chemical potential, Weyl-superconductor phases persist when the Fermi velocities of the Weyl semimetal $(K_0^2\pm 2m\mu_\text{W})^{1/2}/m$ are close to that of the superconductor $Q_0/m'$.
On the other hand, when the Fermi velocities are largely different ($\alpha\ll1$), the phase diagram is dominated by topological superconductor phases [Fig.~\ref{fig:phasediagram_fermilevel03}(d)]. In this case, however, the topological-superconductor phases with even Chern numbers do not shrink as in Fig.~\ref{fig:phasediagram}(d), since we take relatively large chemical potential to make its effect obvious. 
Therefore, we can conclude that, as long as the deviation of the Fermi level away from the node is relatively small ($2m\mu_\text{W}/K_0^2\ll1$), the electronic properties of the multilayer model mentioned in the previous section for the case of $\mu_\text{W}=0$ still hold.
Notice that the surface of the doped Weyl superconductors can host exotic surface states depending on symmetry \cite{lu15}.

Notice also that, in doped Weyl semimetals (not the multilayer model studied in this paper), the possibility of intrinsic superconducting orders by inter-valley pairing (BCS state) and intra-valley pairing (FFLO state) has been studied \cite{cho12, bednik15}.

\section{Conclusion}
\label{sec:conclusion}

We studied electronic properties of a multilayer structure consisting of time-reversal-symmetry-broken Weyl semimetals and s-wave superconductors, in order to examine the superconducting proximity effect in the bulk of Weyl semimetals.
The multilayer model was solved by the method of the Kronig-Penney model.
We first determined the phase diagram from zero-energy solutions, and then showed that the low-energy properties are described by the Weyl Hamiltonian by using the degenerate perturbation theory.

Depending on the thickness of Weyl-semimetal layers and the position of the Weyl nodes, the superconducting proximity effect from superconductor layers splits 2 Weyl nodes into 4 nodes of the Majorana fermions, resulting in Weyl-superconductor phases, or it gaps out to be the three-dimensional extension of topological superconductors, which are fully gapped and characterized by the Chern number. 
In addition, in Weyl-superconductor phases, mismatch of the Fermi velocity of the Weyl semimetal and the superconductor gaps out a pair of Majorana nodes to become Weyl-superconductor phases with 2 Majorana nodes or gaps out all 4 nodes to be topological-superconductor phases that have odd Chern numbers.
Consequently, when the Fermi level of the Weyl semimetal is at the node, the two Fermi velocities match, and the potential barriers are absent, the electronic phase is either a Weyl-semimetal phase with 4 Majorana nodes or topological-superconductor phases with even Chern numbers.
On the other hand, when mismatch of the Fermi velocity becomes larger or when high potential barriers are inserted at the interface, topological-superconductor phases with odd Chern numbers are most likely to be realized.
These features are robust even when the Fermi level is slightly away from the node.

Nontrivial topology of topological-superconductor phases and Weyl-superconductor phases can be detected through the thermal Hall conductivity, which is quantized in topological-superconductor phases and varies continuously by the position of the Majorana nodes in Weyl-superconductor phases.

\acknowledgements
This work was supported by JSPS KAKENHI Grants Nos.~JP17K17604, JP15H05854, and JP17K05485, and JST CREST Grant No.~JPMJCR18T2.

\appendix

\section{Eigenstates of $k_x=k_y=0$ when $\mu_\text{W}=0$}
\label{sec:eigenvaluequation}

For a fixed momenta $k_x, k_y$ and an energy $E$, corresponding eigenfunctions in each layer have eight-fold degeneracy, consisting of four-component spinors of the spin and Nambu space, and two momenta in the $z$ direction of the opposite signs.
The linear combination of these basis functions has 16 coefficients, and they are determined by 4 boundary conditions at $z=0$ and $z=a=-b$ comprising 16 linear equations. The equations are described by a 16 $\times$ 16 matrix.
A nonvanishing solution to $A_{\alpha\beta},B_{\alpha\beta},C_{\alpha\gamma}$ and $D_{\alpha\gamma}$ is present when the determinant of the 16 $\times$ 16 matrix vanishes.

However, the determinant is decomposed into the product of two determinants of 8$ \times$ 8 matrices, when $k_x=k_y=0$.
The vanishing condition of two determinants is given by
\begin{widetext}
\begin{align}
 &
 \left(
 \cos[k_z(a+b)]
 -
 \cos[K_\pm a]\cos[Q_+b]
 +
 \frac{r}{2}\frac{u^2+v^2}{u^2-v^2}
 \left(
 \frac{v_{Q+}}{v_{K\pm}}
 +
 \frac{v_{K\pm}}{v_{Q+}}
 \left(
 1
 +
 \frac{4h^2}{v_{K\pm}^2}
 \right)
 \right)
 \sin[K_\pm a]\sin[Q_+b]
 \right) \notag\\
 &\times
 \left(
 \cos[k_z(a+b)]
 -
 \cos[K_\pm a]\cos[Q_-b]
 -
 \frac{r}{2}\frac{u^2+v^2}{u^2-v^2}
 \left(
 \frac{v_{Q-}}{v_{K\pm}}
 +
 \frac{v_{K\pm}}{v_{Q-}}
 \left(
 1
 +
 \frac{4h^2}{v_{K\pm}^2}
 \right)
 \right)
 \sin[K_\pm a]\sin[Q_-b]
 \right) \notag\\
 &-
 \frac{1}{2}
 \left[
 1-\left(\frac{u^2+v^2}{u^2-v^2}\right)^2
 \right]
 \sin^2[K_\pm a] 
 \left(1-\cos[Q_+b]\cos[Q_-b]\right) \notag\\
 &-
 \frac{1}{4}
 \left[
 1-\left(\frac{u^2+v^2}{u^2-v^2}\right)^2
 \right]
 \left[
 \left(
 \frac{v_{Q+}}{v_{K\pm}}
 +
 \frac{v_{K\pm}}{v_{Q+}}
 \left(
 1
 +
 \frac{4h^2}{v_{K\pm}^2}
 \right)
 \right)
 \left(
 \frac{v_{Q-}}{v_{K\pm}}
 +
 \frac{v_{K\pm}}{v_{Q-}}
 \left(
 1
 +
 \frac{4h^2}{v_{K\pm}^2}
 \right)
 \right)
 -
 \frac{v_{Q+}}{v_{Q-}}
 -
 \frac{v_{Q-}}{v_{Q+}}
 \right]
 \sin^2[K_\pm a]
 \sin[Q_+b]\sin[Q_-b]=0,
 \label{eq:eigenvalueequation}
\end{align}
\end{widetext}
where $v_{K\pm}=K_\pm/m\equiv [K_0^2\pm 2m|E|]^{1/2}/m$, $v_{Q\pm}=Q_\pm/m'$, and $r=1(-1)$ when the first and fourth (second and third) components are relevant.
Reminding that the first and fourth (second and third) components are relevant for modes with $K_+$ when $E>0$ ($E<0$), and with $K_-$ when $E<0$ ($E>0$), $r=1(-1)$ corresponds to the modes with $K_+ (K_-)$ when $E>0$, while $r=1(-1)$ corresponds to $K_- (K_+)$ when $E<0$.
In the limit of the zero energy, $(u^2+v^2)/(u^2-v^2)$ vanishes and, in addition, in the limit of vanishing potential barrier  $h\to 0$, the equation is characterized by a single parameter
\begin{align}
 \frac{v_{Q-}v_{Q+}}{v_{K\pm}^2}
 +
 \frac{v_{K\pm}^2}{v_{Q-}v_{Q+}}.
 \label{eq:mismatch}
\end{align}
This quantity represents mismatch of the Fermi velocity $v_{K_0\pm}=K_0/m$ and $|v_{Q_0\pm}|=|Q_{\pm}(E=0)|/m'=Q_0/m'$. (\ref{eq:mismatch}) gives the minimum value 2 when the two Fermi velocities match, and is greater than 2 when they mismatch.

\section{Low-energy effective Hamiltonian}
\label{sec:degenerateperturbation}

In this section, the degenerate perturbation theory is used to derive a low-energy effective Hamiltonian around a node of Weyl superconductors.

Assume that the chemical potential of the Weyl semimetal $\mu_\text{W}$ is positive.
By taking the limit in order of $k_x,k_y\to 0$ and $E\to +0$, the spinors become
\begin{align}
 &\phi_{+e} 
 =
 \begin{pmatrix}
  1 \\
  0\\
  0\\
  0
 \end{pmatrix},\,
 \phi_{-e} 
 =
 \begin{pmatrix}
  0\\
  1 \\
  0\\
  0
 \end{pmatrix},\,
 \phi_{+h} 
 =
 \begin{pmatrix}
  0\\
  0\\
  1\\
  0
 \end{pmatrix},\,
 \phi_{-h} 
 =
 \begin{pmatrix}
  0\\
  0\\
  0 \\
  1
 \end{pmatrix},
\end{align}
and 
\begin{align}
 &\phi_{+\uparrow}
 \propto
 \begin{pmatrix}
  i\\
  0\\
  0\\
  1
 \end{pmatrix},\,
 \phi_{-\uparrow}
 \propto
 \begin{pmatrix}
  1\\
  0\\
  0\\
  i
 \end{pmatrix},\,
 \phi_{+\downarrow}
 \propto
 \begin{pmatrix}
  0\\
  1\\
  i\\
  0
 \end{pmatrix},\,
 \phi_{-\downarrow}
 \propto
 \begin{pmatrix}
  0\\
  i\\
  1\\
  0
 \end{pmatrix}.
\end{align}
Two eigenfunctions at a node $\bm{k}^{(0)}=(0,0,k_z^{0})$ given by
\begin{align}
 \psi_1(z)
 &=
 \left\{
 \begin{array}{ll}
  \displaystyle
   \phi_{+e} \chi_{+e}(z)+
   \phi_{-h} \chi_{-h}(z)
  \,\, &(z\in(0,a))\\
  \displaystyle
  \sum_{\alpha=\pm}
   \phi_{\alpha\uparrow}\chi_{\alpha\uparrow}(z)
  \,\, &(z\in(-b,0))
 \end{array}
 \right.
\end{align}
and
\begin{align}
 \psi_2(z)
 &=
 \left\{
 \begin{array}{ll}
  \displaystyle
   \phi_{-e} \chi_{-e}(z)+
   \phi_{+h} \chi_{+h}(z)
  \,\, &(z\in(0,a))\\
  \displaystyle
  \sum_{\alpha=\pm}
   \phi_{\alpha\downarrow}\chi_{\alpha\downarrow}(z)
  \,\, &(z\in(-b,0))
 \end{array}
 \right.
\end{align}
satisfy $H(0,0,-i\partial_z)\psi_i(z)=0\,(i=1,2)$, where the wave function in a Weyl-semimetal layer $z\in(0,a)$ for a spinor $\phi_{\alpha\beta}$ is denoted by $\chi_{\alpha\beta}(z)=A_{\alpha\beta}e^{iK_{\alpha}z}+B_{\alpha\beta}e^{-iK_{\alpha}z}\,(K_{\alpha=\pm}^2=K_0^2\pm 2m|\mu_\text{W}|)$, and that in a superconductor layer $z\in(-b,0)$ for a spinor $\phi_{\alpha\gamma}$ by $\chi_{\alpha\gamma}(z)=C_{\alpha\gamma}e^{iQ_{\alpha}z}+D_{\alpha\gamma}e^{-iQ_{\alpha}z}$ [$Q_{\alpha=\pm}^2=2m'(\mu_\text{S}\pm i|\Delta|)$].
Up to the first order in the deviation of the momentum $\bm{p}=\bm{k}-\bm{k}^{(0)}$, the eigenfunction is a linear combination of the 0-th order functions given by $e^{ip_z z}(C_1\psi_1(z)+C_2\psi_2(z))$.
The perturbation Hamiltonian is
\begin{align}
 &e^{-ip_z z}H(p_x,p_y,-i\partial_z)e^{ip_z z}
 -
 H(0,0,-i\partial_z) \notag\\
 =
 &H(p_x,p_y,-i\partial_z+p_z)
 -
 H(0,0,-i\partial_z)\notag\\
 \simeq
 &\left\{
 \begin{array}{ll}
  p_z\left(-i \partial_z\right)\sigma^z\tau^z/m
  + \lambda(p_y\sigma^x-p_x\sigma^y\tau^z)\,\, &(z\in(0,a))\\[+6pt]
  p_z\left(-i \partial_z\right)\tau^z/m' \quad &(z\in(-b,0))
 \end{array}
 \right..
\end{align}
Then, the effective Hamiltonian is given by
\begin{align}
 H_\text{eff}^{ij}(\bm{p})
 &\equiv 
 \frac{1}{\left(
 c_ic_j
 \right)^{1/2}}
 \int_{-b}^{a}dz
 \psi_{i}^{\dagger}
 H(p_x,p_y,-i\partial_z+p_z)
 \psi_j \notag\\
 &=
 \begin{pmatrix}
  v_1p_z 
  & v_2(p_y+ip_x) \\
  v_2^{\ast}(p_y-ip_x) 
  & -\tilde{v}_1p_z
 \end{pmatrix}_{ij},
 \label{eq:effectivehamiltonian}
\end{align} 
where $v_1=(v_{+}+v_\uparrow)/c_1$, $\tilde{v}_1=(v_{-}+v_\downarrow)/c_2$, $v_2=\lambda w/(c_1c_2)^{1/2}$, and
\begin{align}
 &v_{\pm}
 =
 \frac{1}{m}\int_0^adz
 \left[
 \chi_{\pm e}^{\dagger}(-i\partial_z)\chi_{\pm e}
 +
 \chi_{\mp h}^{\dagger}(-i\partial_z)\chi_{\mp h}
 \right], \\
 &v_{\gamma}
 =
 \frac{1}{m'}
 \int_{-b}^0dz
 \left[
 i\chi_{-\gamma}^{\dagger}(-i\partial_z)\chi_{+\gamma}
 -
 i\chi_{+\gamma}^{\dagger}(-i\partial_z)\chi_{-\gamma}
 \right]\,(\gamma=\uparrow,\downarrow), \\
 &w
 =
 \int_0^adz
 \left[
 \chi_{+e}^{\dagger}\chi_{-e}
 +
 \chi_{-h}^{\dagger}\chi_{+h}
 \right].
\end{align}
Here, $c_i$ is the normalization constant defined by
\begin{align}
 c_i
 =
 \int_{-b}^{a}dz
 \left|
 \psi_i
 \right|^2.
\end{align}
From equations for $A_{\alpha\beta},B_{\alpha\beta},C_{\alpha\gamma}$ and $D_{\alpha\gamma}$, identities
$\chi_{+e}=\chi_{+h}$, $\chi_{-e}=\chi_{-h}$, $\chi_{+\uparrow}=\chi_{+\downarrow}$, and $\chi_{-\uparrow}=\chi_{-\downarrow}$ hold up to an irrelevant phase factor when $E=0$ and $k_x=k_y=0$.
This leads to relations $v_+=v_-$, $v_{\uparrow}=v_{\downarrow}$, $w\in\mathbb{R}$, $c_1=c_2$, and thus $v_1=\tilde{v}_1$, $v_2^{\ast}=v_2$.
The relation $v_1=\tilde{v}_1$ is consistent with symmetries, that is, eigenenergies come with positive- and negative-energy pairs.

\begin{figure}
 \centering
 \begin{tabular}{cc}
  \begin{minipage}[c]{40mm}
   \includegraphics[width=40mm]{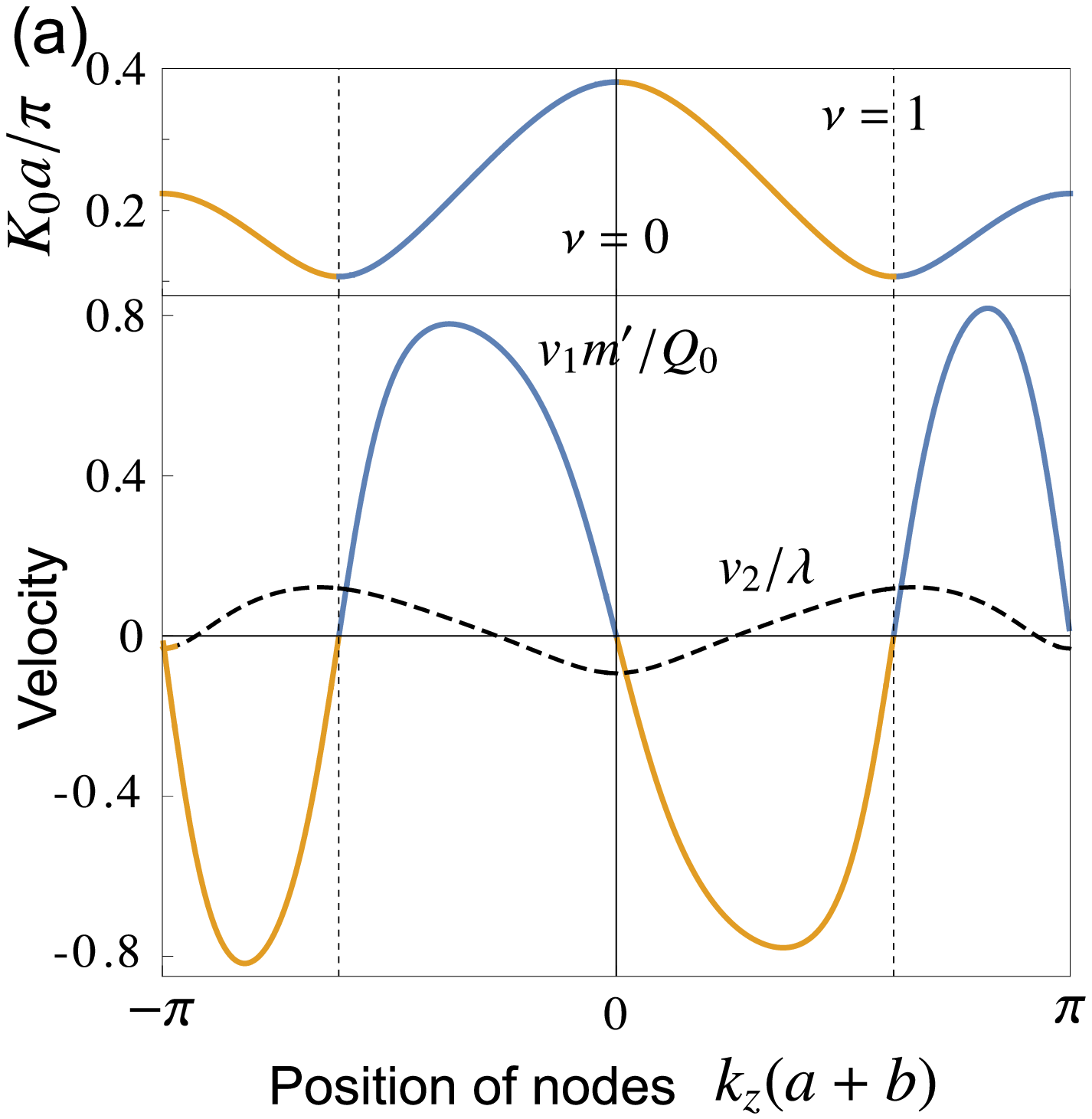}
  \end{minipage}&
  \begin{minipage}[c]{40mm}
   \includegraphics[width=40mm]{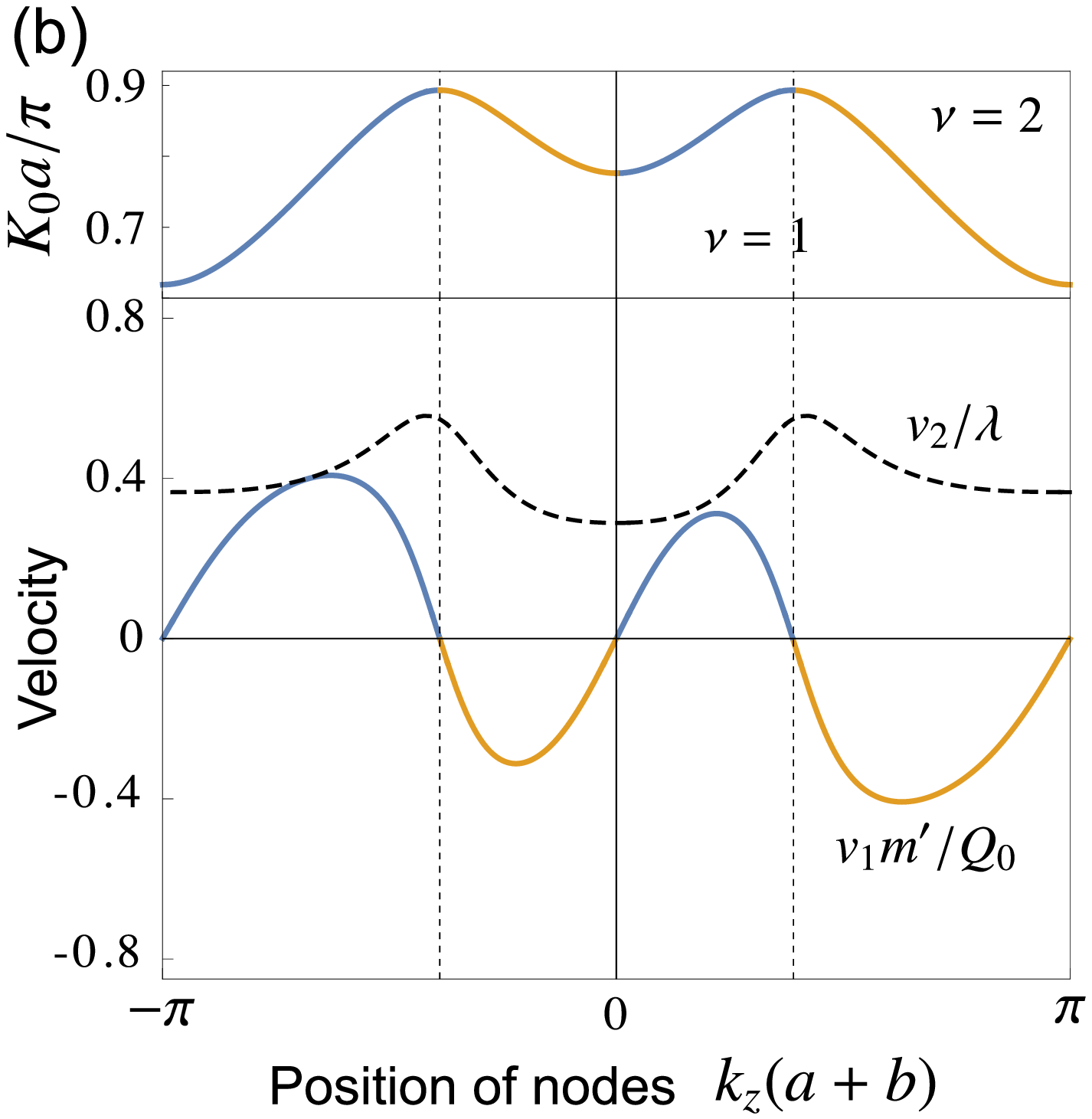}
  \end{minipage}
 \end{tabular}\\
 \begin{tabular}{cc}
  \begin{minipage}[c]{40mm}
   \includegraphics[width=40mm]{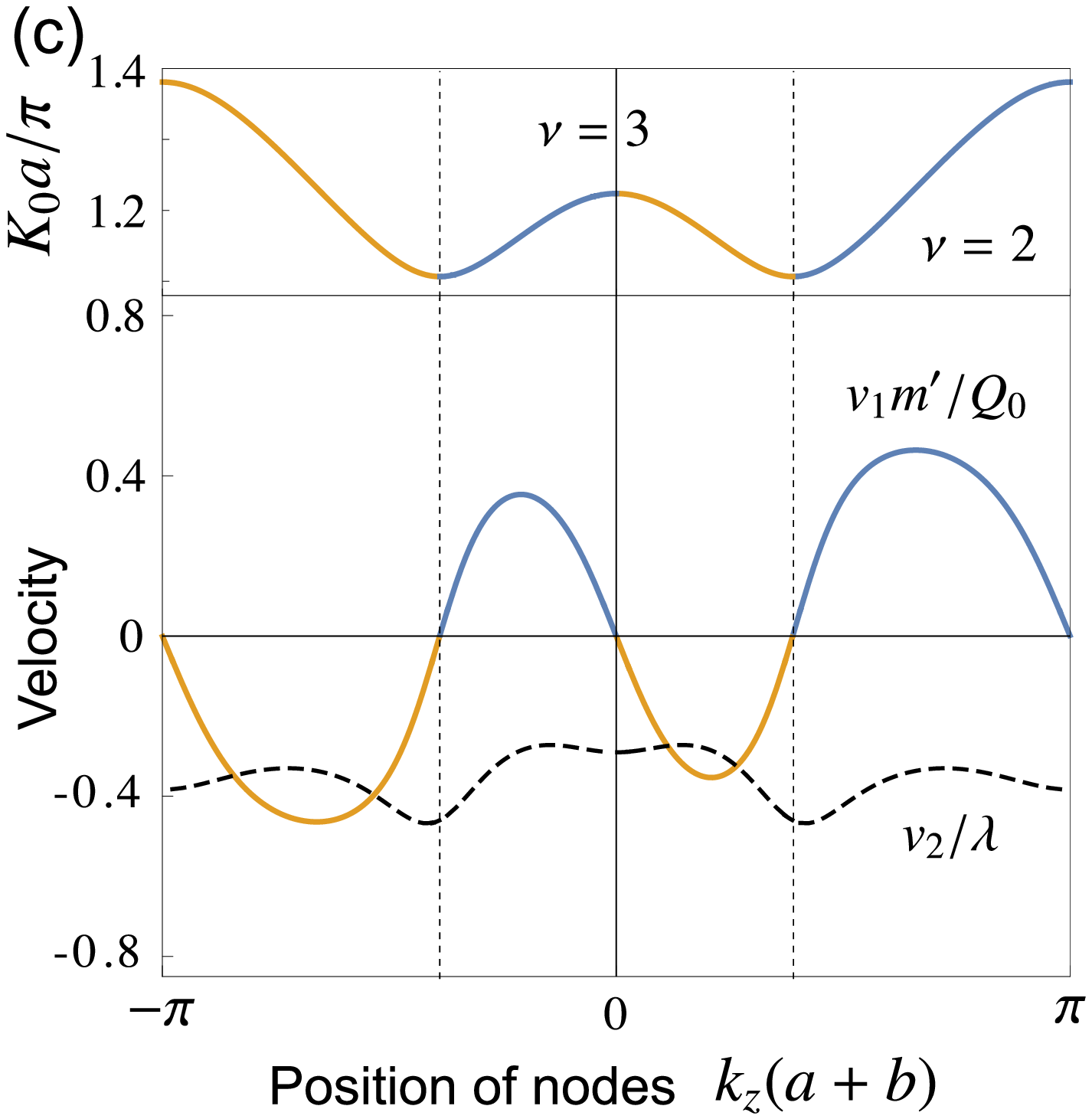}
  \end{minipage}&
  \begin{minipage}[c]{40mm}
   \includegraphics[width=40mm]{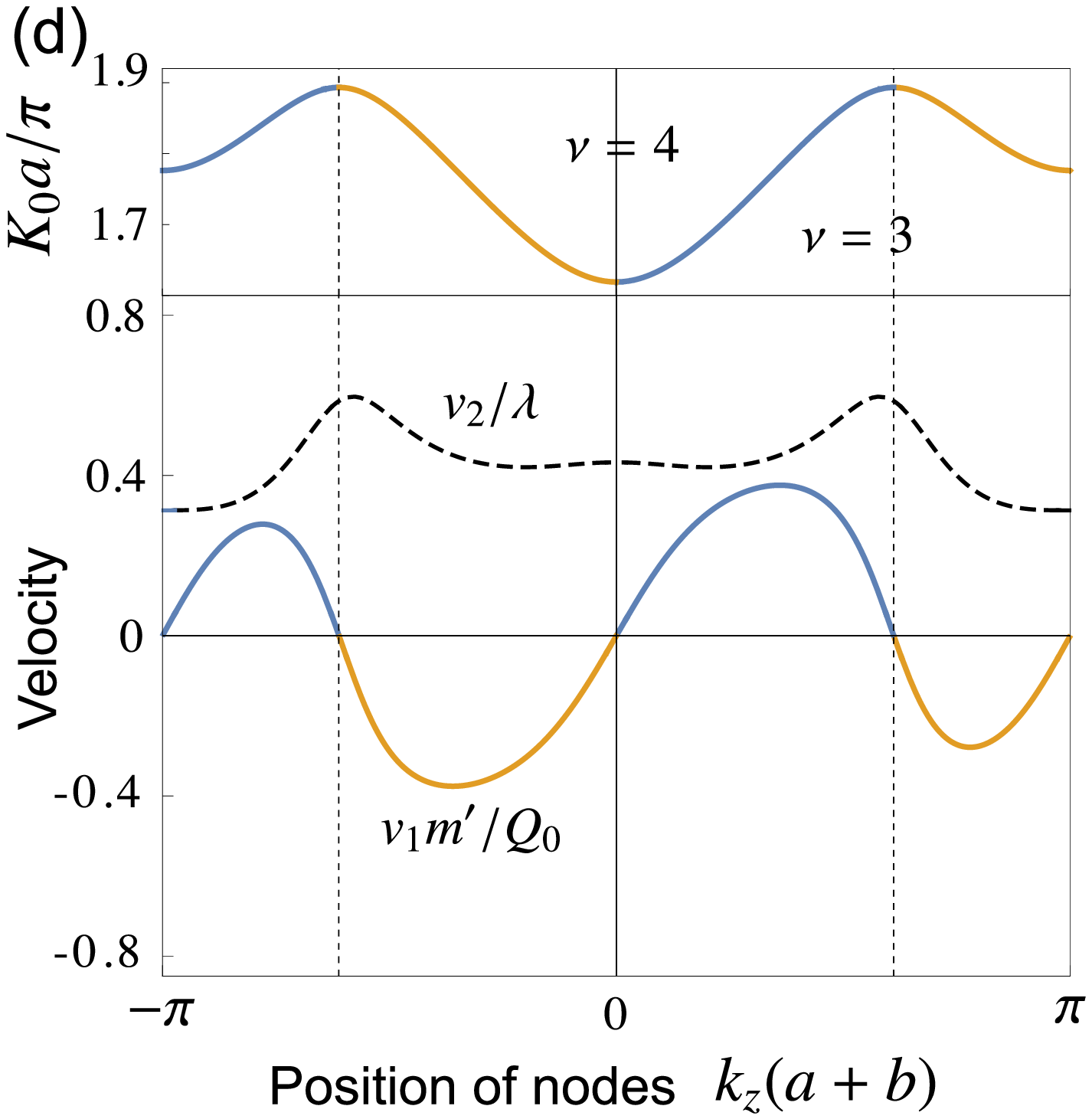}
  \end{minipage}
 \end{tabular}
 \caption{(Color online) The coefficients $v_1$ (solid line) and $v_2$ (dashed line) in the effective Hamiltonian (\ref{eq:effectivehamiltonian}) normalized by the Fermi velocity $Q_0/m'$ and Rashba coupling $\lambda$, respectively, are shown as a function of the position of nodes. Parameters $\mu_\text{W}\to +0$, $\alpha=(K_0/m)/(Q_0/m')=1/2$, $Q_{0}b/\pi=1.4$ are used, which are the same as those in Fig.~\ref{fig:positionofweylnodes}(a). In addition, $m'/m=1$ is used. Each figure corresponds to (a) the lowest line, (b) the second-, (c) third-, and (d) fourth-lowest line in Fig.~\ref{fig:positionofweylnodes}(a), respectively. The position of the node is also shown on the top as a function of $K_0a$. Blue (yellow) curves correspond to positive (negative) chirality nodes. Vertical dotted lines represent momenta where two nodes merge besides the center and the boundary of the Brillouin zone ($k_z(a+b)=0, \pi$).
 \label{fig:effectivehamiltoniancoefficients}}
\end{figure}
In Fig.~\ref{fig:effectivehamiltoniancoefficients}, the coefficients $v_1$ and $v_2$ in the effective Hamiltonian (\ref{eq:effectivehamiltonian}) are numerical evaluated for $\mu_\text{W}\to +0$, $\alpha=(K_0/m)/(Q_0/m')=1/2$, $Q_0b/\pi=1.4$, and $m'/m=1$. Fig.~\ref{fig:effectivehamiltoniancoefficients}(a) [(b), (c), (d)] corresponds to $K_0a/\pi\in [0,0.5]\, ([0.5,1], [1,1.5], [1.5,2])$, in which $v_1$ and $v_2$ are evaluated at the node shown in the lowest (second, third, fourth lowest) line of Fig.~\ref{fig:positionofweylnodes}(a) separating $\nu=0$ (1, 2, 3) and $\nu=1 (2, 3, 4)$ regions. 
The sign of a velocity $v_1$ determines the monopole charge of the node. On the other hand, the sign of a velocity $v_2$ is irrelevant to the monopole charge, while $v_2$ must be nonvanishing to guarantee the linear dispersion in the $k_x$ and $k_y$ direction. It can be seen from Fig.~\ref{fig:effectivehamiltoniancoefficients} that a pair of nodes that emerge from or annihilate at the same momentum have the opposite monopole charge.

\end{document}